\documentclass[aps,pra,twocolumn,superscriptaddress,nobibnotes,a4paper,10pt]{revtex4-2}
\pdfoutput=1

\usepackage{graphicx}
\usepackage{physics}
\usepackage{siunitx}
\usepackage{dcolumn}
\usepackage{bm}
\usepackage[colorlinks=true,linkcolor=blue,urlcolor=blue,citecolor=blue]{hyperref}
\usepackage{natbib}
\usepackage[table]{xcolor}
 \usepackage{setspace}
\usepackage{makecell}

\definecolor{lightblue}{RGB}{184,213,216}
\definecolor{lightyellow}{RGB}{247,223,169}

\begin{document}

\title{Scalable Quantum Computing with Optical Links}

\author{M.J.~Weaver}\email{matthew@qphox.eu}
\author{G.~Arnold}
\author{H.~Weaver}
\author{S.~Gr\"oblacher}\email{simon@qphox.eu}
\author{R.~Stockill}\email{rob@qphox.eu}
\affiliation{QphoX B.V., Elektronicaweg 10, 2628XG, Delft, The Netherlands}

\begin{abstract}
Quantum computers have great potential to solve problems which are intractable on classical computers. However, quantum processors have not yet reached the required scale to run applications which outperform traditional computers. Leading hardware platforms, such as superconducting qubit based processors, will soon become bottlenecked by the physical constraints of their low temperature environments, and the expansion of quantum computers will necessitate quantum links between multiple processor modules. Optical frequencies offer the most promising path for these links due to their resilience to noise even at ambient temperature and the maturity of classical optical networks. However, required microwave-to-optics transducers cannot operate deterministically yet, which has widely been seen as a key challenge for their integration into fault-tolerant quantum computers. In this work, we examine implementations of optical links between cryogenic units that surpass the performance of individual cryogenic modules even with the performance of existing or near-term microwave-to-optics transducers. We show methods for these transducers to provide on-demand entanglement between separated quantum processors with high fidelity and lay out key steps for adoption of the technology including scaling transducer numbers and integration with other hardware. Finally, we discuss a number of architectures comprised of these links which can drive the expansion of quantum data centers to utility scale.
\end{abstract}
\maketitle

\tableofcontents
\newpage

\section{Introduction}
The past decade has seen a tremendous surge in efforts to build a first viable quantum computer, with the goal to revolutionize computing and deliver on the potential of quantum mechanics for advanced real world applications~\cite{Horowitz2019,Alexeev2021,Gyongyosi2019}. The experimental and technical focus has been very broad with teams pursuing many different physical systems, ranging from ions~\cite{Bruzewicz2019,Strohm2024}, photons~\cite{Slussarenko2019}, superconducting circuits~\cite{Kjaergaard2020,Bravyi2022} to neutral atoms~\cite{Henriet2020,Wintersperger2023}, to name a few. While some approaches, such as superconducting qubits scale favorably via nanofabrication techniques~\cite{Bravyi2022,Gold2021,Arute2019}, they are often limited by the qubit gate fidelity and therefore require complex error correction algorithms~\cite{Roffe2019,Acharya2025}. Other realizations, such as trapped ions, have demonstrated higher gate fidelities, however, currently lack a clear path for scaling to large enough qubit numbers within individual processors~\cite{Strohm2024}. Even the most ambitious roadmaps for building a universal quantum computer predict that it will still take at least a decade of research \& development before quantum computers will become useful tools for advancing materials science, medicine, communications and other impactful fields~\cite{Dalzell2023}. For all quantum computing (QC) systems, significant uncertainty and technical challenges remain.

In order to realize mature quantum computers, applying strategies from classical computers and data centers could significantly accelerate the implementation of solutions to current challenges~\cite{Cheng2018,Baziana2024}. Building modular clusters of interlinked quantum processors could effectively increase the qubit count with existing QC technologies, avoiding the hard technical limitations to quantum processor size~\cite{Barral2025,Awschalom2021,Caleffi2024,Boschero2024}. Additionally, combining the strengths of individual platforms and therefore overcoming their limitations has also been proposed~\cite{Caleffi2024}. These approaches require developing an interface through a common network frequency which can freely exchange quantum information~\cite{Awschalom2021}. The quantum links that result between the processors, which are comprised of non-local entangled states, need to be low noise, high fidelity and on-demand in order to seamlessly connect processors into a large scale quantum computer.

Realizing such quantum links has been the focus of intense research over the past years, where quantum information is converted to a suitable carrier, such as optical photons, in order to distribute entanglement between physically separated qubits~\cite{Moehring2007,Ritter2012,Bernien2013}. While this is not the only possible approach~\cite{Magnard2020, Devitt2016}, optical photons have many distinct advantages, such as a high resilience to noise and low-loss propagation, especially in the telecommunications frequency band~\cite{Kanamori2021}, and are therefore the preferred carrier of quantum information when linking individual quantum systems together. In order to reach a suitable interfacing wavelength, such as the optical telecom band, some of the most advanced implementations of quantum frequency conversion (QFC) rely on either non-linear optics for visible to telecom optics conversion~\cite{Rakher2010} and for microwave-to-optics conversion~\cite{Han2021}, or transfer through an intermediary system, such as a mechanical resonator~\cite{Chu2020,Lambert2020}. Current technology for converting quantum states from microwave frequencies to optical frequenices, often also referred to as quantum transduction, has been able to demonstrate efficiencies approaching 50\%~\cite{Brubaker2022}, however with significant limitations in bandwidth and added noise remaining. While there is no fundamental limitation preventing entanglement between remote qubits at high repetition rates, several technical challenges remain.

Due to the physical limitations of various quantum systems being explored for quantum processing, memories and sensing, some form of QFC will almost certainly be required for either scaling quantum computing and/or networking with other quantum systems. However, quantum transduction is often viewed as only practical when reaching performance levels that approach deterministic entanglement generation, including near unity efficiency, MHz generation rates and negligible added noise, which are far from the current state-of-the-art. In this article, we challenge this prevailing viewpoint and highlight possible protocols using transducers with existing or near term performance levels, which could achieve on-demand links with greater than 99$\%$ Bell-state fidelity. We discuss how despite their imperfect operation, they can already now boost scalability and performance levels of existing quantum processors. We will discuss the minimum requirements for transducers to have real impact and how the transducers allow for entanglement generation between qubits which will result in new opportunities and changes to quantum processor operations. Our vision will directly lead to the first use cases of quantum transducers for QC. Throughout this overview we will focus mostly on superconducting qubits in combination with microwave-to-optics transduction in order to determine the parameter space available in the near term as an illustrative example. The same ideas and principles, however, also apply to other frequency domains, such as quantum systems in the visible domain.

\begin{figure}
\includegraphics[width=0.9\linewidth]{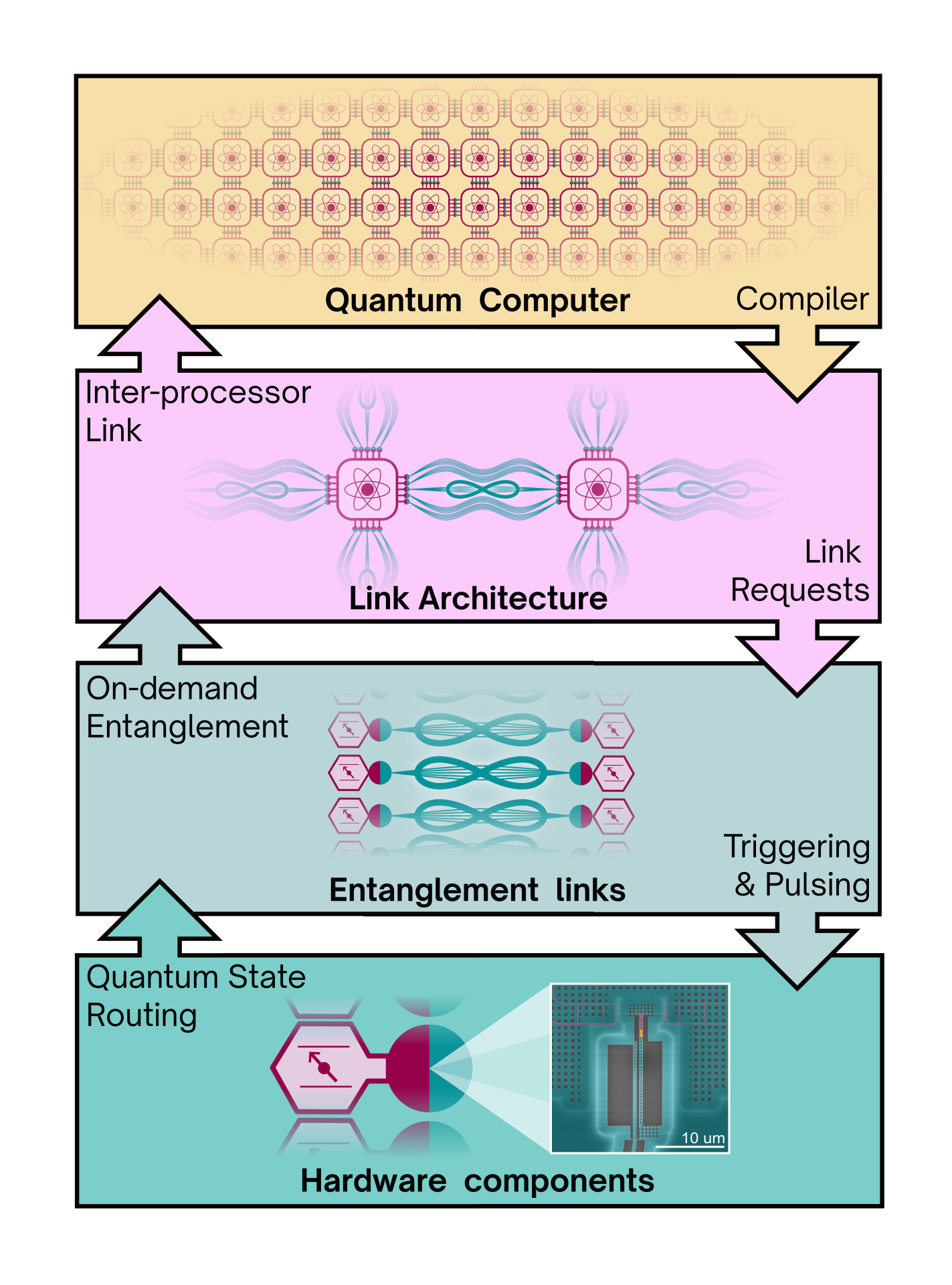}
\caption{Overview of the quantum computer hardware stack in a distributed system. At the base layer there are hardware components such as qubits and microwave-to-optics transducers, which are connected to each other and coordinated via electrical or optical pulses.  The hardware components route quantum states and together comprise the second level: entanglement links. Multiple entanglement links can be requested and delivered (on-demand) to generate intermodule links which match the requirements for distributed gates. The link architecture can be dynamically adapted by a compiler to provide the necessary inter-processor links to perform the quantum computation.}
\label{fig:overview}
\end{figure}

A number of works have expored the requirements for a distributed quantum computer and categorized the constituent technologies into a layered network stack ~\cite{Barral2025,Caleffi2024,Boschero2024,Wehner2018,Ang2024}. Developments in several fields are required including algorithms, software, firmware and hardware~\cite{Caleffi2024,Wehner2018,Parekh2021,Ferrari2021}. In this article we focus on the hardware, which spans the physical layer and the link layer of the network stack~\cite{Wehner2018,Ang2024}. Figure ~\ref{fig:overview} shows an overview of the hardware stack for a distributed quantum computer. The quantum computer is composed of modular quantum processors which are connected with inter-processor links. The inter-processor links are broken into a hardware components layer, and an entanglement link layer. By using on-demand entanglement links as a building block for inter-processor links, quantum computers can deterministically run computations with remote processors as resources, which simplifies compilation.

We begin in section \ref{section:figuresofmerit} with a review of some of the most important parameters to consider for effective operation of quantum frequency converters. We then discuss the performance of state-of-the-art transducers in achieving these metrics and how with modest improvements transducers can achieve the necessary performance to serve as building blocks for quantum links (section \ref{section:transducerreview}). Having established realistic estimates of transducer performance we present a number of exemplary protocols and configurations for generating on-demand entanglement links with high fidelity which overcome any limits to transducer efficiency and noise (section \ref{section:quantumlinks}). Finally, in section \ref{section:architecture} we explore some architectures and link configurations which could provide a near-term benefit to quantum computers. Scaling up transducer technology to meet the requirements of these near-term optical links will accelerate this foundational technology towards eventual full-capacity fault-tolerant optical links between quantum processors.

\section{Figures of merit}
\label{section:figuresofmerit}

Remote entanglement of qubits via an optical link, especially microwave frequency qubits, remains a demanding task. A wide variety of parameters must be simultaneously optimized to faithfully transmit signals between processors. A number of publications have explored the important parameters which must be considered in order to contruct a useful quantum link based on microwave-to-optics transducers~\cite{Zeuthen2020, Han2021}. Here we give a high level summary of the parameters with important additions related to practical operation in the vicinity of a quantum processor.

\subsection{Efficiency}
This quantity ($\eta$) refers to the ratio of the number of optical photons output to microwave photons input or vice versa. It is important to note that all transduction technologies use an optical pump tone to overcome the frequency (and energy) difference between the microwave and optical signals. The efficiency is generally increased with increasing optical pump power, which in turn can introduce more absorption-induced noise to the output signal.

\subsubsection{Fixed Efficiency}
This quantity covers any constant losses in the transducer. Common sources of loss include outcoupling of resonators to external transmission lines, signal reflection due to mismatched state transfer rates between internal modes, intrinsic dissipation from internal modes and lossy connections to optical fibers and microwave waveguides. Collectively these set the maximum achievable efficiency that the transducer can reach.

\subsubsection{Variable Efficiency}
The requirement of an optical-frequency pump to overcome the energy difference between microwave and optical modes and enable the transduction process results in a pump power-dependent transducer efficiency. There is a general trade-off between higher efficiency and higher noise levels due to absorption or scattering from the pump tone. Additional sources of variable efficiency include microwave frequency pumps for double parametric transducers~\cite{Andrews2014} and the additional requirements some realisations have on bias voltages and currents. Understanding the device-specific trade-offs in efficiency, noise and bandwidth that result from these settings and how they relate to performance of different entanglement-distribution protocols are essential for maximising the utility of each quantum transducer.

\subsection{Bandwidth}
Achieving near-unity photon transduction efficiency has generally required enhancing interactions through the use of resonant modes. As a result, the bandwidth of the transducer becomes limited and plays an important role in assessing the efficiency with which it can transfer a photon wavepacket. A particular advantage of superconducting circuits is the speed of operation, and a transducer that can support this must have a sufficiently large bandwidth to enable efficient transduction of the wavepackets. In the case where the transducer is used as a two-mode squeezing source, the bandwidth must be matched to a wavepacket which can be efficiently captured by the qubit or quantum memory.

\subsection{Noise}
This metric refers to the additional noise introduced in the transduction process, and must be particularly small to enable transduction of a quantum state. There are two main noise sources: thermal noise and vacuum noise (see below). Noise is perhaps the principal metric for understanding the potential of a transduction implementation. The heralded entanglement distribution protocols which we will discuss in section \ref{section:quantumlinks} are tolerant to low efficiency transduction. However, any noise approaching the single photon level will corrupt the herald and directly result in lower-fidelity entanglement.

It is important to note that for photon transduction the relevant metric is input-referred noise, which results in differing values for up- and down-conversion. Due to the low thermal occupation of optical-frequency modes at room temperature and below, most noises sources are added at microwave frequencies. In the common case of small optical conversion rates to the optical domain (resulting from small optical cooperativities), optical inputs are typically attenuated more than microwave inputs before noise is added, and hence upconversion from microwave-to-optical frequency generally benefits from much smaller input-referred noise levels. As a result, many entanglement protocols use this technique in combination with very low noise single photon measurement, rather than combining up- and down-conversion in an end-to-end state transfer scheme.

Naturally a different metric (for example noise in an intermediate mode) is required for the added noise if the transducer is used as a two-mode-squeezing source, however the same relevant noise sources must be taken into account.

\subsubsection{Thermal Noise}
A prerequisite for quantum transduction of single photon states is ground state occupation of the transducer modes, necessitating either thermalisation to the mixing chamber of a dilution refrigerator or radiative cooling to a cold bath. 

Enabling ground state occupation via a dilution refrigerator then imposes strict conditions on the power dissipation from the transducer optical pump, due to the limited cooling capacity at millikelvin temperatures. Capping the dissipated power then results in limitations for the efficiency, which depends on the optical pump power, and the maximum repetition rate for the transducer.

For many transduction realisations, local heating is the dominant source of noise. While the total power dissipation is small enough to keep the dilution refrigerator at its base temperature, absorption of the optical pump field results in local heating of the transducer device and raises the thermal occupation of the microwave-frequency modes. This effect can additionally overwhelm any radiative cooling to the optical or microwave bath.

Finally, for transducers that rely on low frequency ($< $GHz) modes, radiative cooling to a higher frequency cold-bath is the only option to realise ground-state occupation. The requirements for efficient radiative cooling match the requirements for transduction, however, can result in additional operational overhead and per-device power dissipation.

\subsubsection{Vacuum Noise}

Optical detectors can also add noise: both quantum limited and classical. If standard optical detectors are used, amplitude fluctuations from shot noise are added, which can be a significant contributor to the added noise if transduction efficiency is low. The quantum channel capacity is 0 for transduction efficiencies below 50$\%$~\cite{Zhong2022}. However, single photon detectors have much lower noise than the shot noise level at the cost of loss of phase information. With these detectors and a classical communication channel, quantum information can be transferred at far lower efficiencies (see Section \ref{section:quantumlinks}). Classical noise is also added in the form of voltage fluctuations or dark counts on the detectors. For the transducer, an additional source of vacuum noise results from spontaneous splitting of an optical pump photon into a microwave-frequency signal and a lower-frequency photon. This effect can be suppressed with the pump detuning from the optical resonator, which sets the requirement that the optical resonator must have a loss rate smaller than the frequency of the low-frequency mode (often referred to as the sideband resolved regime).

\subsection{Integration Metrics}
Beyond the core metrics which describe the performance of a standalone microwave-to-optics transducer, a number of practical metrics need to be taken into account to understand how large numbers of channels can be simultaneously successfully operated to collectively realise the high-rate high-fidelity entanglement required to link quantum processor modules together.

\subsubsection{Power Dissipation}
In order to scale to many transduction channels operating simultaneously while keeping the cryostat at base temperature, the optical power required to enable efficient transduction must be minimised. While in principle the dissipation of any optical pump can be reduced through efficient coupling and minimising intrinsic loss of transducer modes, operating tens or hundreds of transducer channels can easily overwhelm the sub-mW cooling that is generally possible at millikelvin temperatures. In order to encapsulate this the metric of efficiency/ pump power (typically \%/$\mu$W) is employed.

\subsubsection{Repetition Rate}
A common approach to minimise the adverse effects of the optical pump is through pulsing the transducer operation. This ensures that power dissipation is only occurring while the transducer is operational, and potentially that transduction occurs before the onset of any increased thermal noise. While compatible with on-demand operation, pulsing this way has an associated maximum repetition rate for the transducer to return to its ground state. The rate is ultimately capped at the transduction bandwidth (for continuous transducer operation).

\subsubsection{Integration Requirements}
A number of other factors determine the ability to scale up the use of a transducer to the point where the system can successfully link remote quantum processors. These include the ability to achieve chip-scale multiplexing of signals and simultaneous manufacture of transducer channels, the ability to reproduce the operating frequencies to achieve optical indistinguishability and the footprint required to accommodate each transducer channel. Additionally, the transducers must also operate at the frequency of the quantum system, either by design, or, preferably through some local tunability.

The additional requirements for operating a particular transducer realisation beyond the optical pump, be it a magnetic or electric field, bias voltage or current or microwave-frequency pump must be considered, and how these fields can be isolated from the quantum processor. Finally, the requirements for the external, room-temperature hardware must be considered, including the feasibility of locking and then filtering the optical pump light for the number of channels required, and the compatibility with the low-loss telecom wavelengths around 1550 nm.

\subsection{Emergent Metrics}
Taken together the hardware metrics for a single transducer realization can be reinterpreted in terms of the performance such a configuration can provide, such as the maximum fidelity of an entangled state that could be generated between remote qubits, and the rate at which such a state could be realised. Scaling to multiple copies of the transducer, one can then consider the fidelity of on-demand entanglement, and with multiple entangled state copies, the fidelity and overhead of a purified entangled state. Finally, taken on a larger scale, one can examine broader metrics such as the quantum volume or maximum circuit depth that could be achieved with a distributed architecture realised through a particular transducer. At each level one is required to make additional selections of either  the correct entanglement distribution protocol (section \ref{section:quantumlinks}) or the global archictecture for the distributed quantum computer (Section \ref{section:architecture}).

\section{State-of-the-art in quantum transduction}
\label{section:transducerreview}

In this section we will discuss some of the advances in microwave-to-optics transducer research. A number of excellent reviews already exist on this subject and give an overview of the extreme breadth of approaches which are being explored~\cite{Han2021,Kurizki2015,Lambert2020,Lauk2020,Chu2020,Clerk2020}. Here we will constrain ourselves to a narrow subset of transducers by focusing first on the integration criteria. 

As we will see in Sections \ref{section:quantumlinks} and \ref{section:architecture}, to fully utilize the power of transduction in a quantum computer it is likely that many links per quantum processor will be required. As a result, a large number of transducers will need to fit on the mixing chamber of a dilution refrigerator, without producing too much heat. Furthermore, these transducers will need interfaces to state-of-the-art large-scale processors.  For these reasons, transducer devices which are integrated on a chip, with high channel density provide a practical approach to scaling. So far electro-optic transducers, electro-opto-mechanical transducers and rare earth ion transducers have been integrated onto a chip.

\begin{table*}[ht]
    \centering
    \caption{Examples of transducers. *Repetition time is limited by the bandwidth. $^\dagger$Microwave-output-referred added noise (Optical-input referred added noise) }
    \begin{tabular}{ccccccc}
Reference & Type &$\eta_{tot}$ &$t_{rep} (\mu s)$ & BW (MHz) & $N_{add}$ & $\eta/p_{o} ($\% $/\mu W)$ \\ \hline
Weaver(2024)\cite{Weaver2024} & EMO & $3\cdot10^{-6}$ & 10 & 15 & 6  & 0.05, 1(CW)\\ 
Jiang(2023)\cite{Jiang2023} & EMO & $1\cdot10^{-4}$ & 5.9 & 1.5  & 2 & \\ 
Brubaker(2022)\cite{Brubaker2022} & EMO &  0.38 & 5,000* & $2.2\cdot10^{-4}$ & 3.2 & 16 \\ 
Meesala(2024)\cite{Meesala2024} & EMO &  $<6\cdot10^{-3}$  & 20 & $\sim$5.5  & 0.14 & \\ 
Zhao(2024)\cite{Zhao2024} & EMO  & $<8\cdot10^{-3}$  & 11.2* & $8.9\cdot10^{-2}$ & 0.94 & 5\\ 
Warner(2025)\cite{Warner2025} & EO & $<1\cdot10^{-3}$ & 1 & 30 & 0.12 (12)$^\dagger$ & 0.05 \\ 
Shen(2024)\cite{Shen2024} & EO & $\ll10^{-4}$  & $6\cdot10^{-5}$* & 17,000 & 23 & $1\cdot10^{-7}$\\ 
Xie(2025)\cite{Xie2025} & REI & $3.4\cdot10^{-5}$ & 10,000 & 0.5 & 1.24 & $1\cdot10^{-5}$ \\ \hline
    \end{tabular} 
    \label{tab:transducersachieved}
\end{table*}

\subsection{Electro-optic Transducers}
Electro-optic transducers modulate optical signals with microwave-frequency electric fields by leveraging the Pockels effect in materials such as lithium-niobate, lithium tantalate, aluminum nitride and, more recently, barium and strontium titanate~\cite{Xu2021,Fan2018,Wang2024a,Mohl2025,Anderson2025}. These devices enable direct photon conversion between the microwave and optical domains, albeit with limited electro-optic coupling rates. To enhance the overall efficiency, microwave and optical resonators are used to increase the interaction time~\cite{Tsang2010}, along with strong optical pumps for parametric enhancement of the electro-optic coupling rate, $G_{EO}\propto\sqrt{P_{opt}}$. As a consequence of the direct coupling principle, optimal efficiency occurs when the pump-photon enhanced coupling rate matches the output coupling rates of the device, defined by the microwave ($\kappa_{MW}$) and optical ($\kappa_{O}$) resonator linewidths. This condition corresponds to an electro-optic cooperativity of $C_{EO}=\frac{4G_{EO}^2}{\kappa_{MW}\kappa_{O}} \approx 1$.

Lithium niobate has emerged as a leading material for electro-optic transducers due to its high electro-optic coefficient and favorable optical properties. Recent integrated electro-optic transducers (central device dimensions $\simeq$$1.5\times0.5$~mm$^2$) have achieved conversion efficiencies around 0.05 \%/$ \mu W$ of optical pump power over a bandwidth of $\sim 30$ MHz with repetition rates of hundreds of kHz~\cite{Warner2025}. The device has been used to send control pulses to a superconducting qubit. Three-dimensional, macroscopic electro-optic transducers have demonstrated entanglement between microwave and optical fields \cite{Sahu2023} and all-optical single shot readout of a superconducting transmon qubit \cite{Arnold2025} with similar conversion-efficiencies as the integrated approach mentioned above, which indicates that entanglement based applications with an interface to qubits may be in reach for scalable devices. 

In a complementary traveling-wave approach, the efficiency-enhancing but bandwidth-limiting optical and microwave resonators were replaced by decimeter-long optical waveguides and surrounding superconducting electrodes, enabling an electro-optic modulation length of 1 meter with a compact footprint of $14\times4$~mm$^2$~\cite{Shen2024}. This integrated traveling-wave microwave-optic transducer demonstrated bandwidths of up to 17 GHz, surpassing even the bandwidths of conventional microwave components. However, optical absorption limited the conversion efficiency of these devices to $10^{-7}$ \%/$ \mu W$ of optical power and required operation at 4K due to the higher cooling power, currently preventing quantum applications with such transducers and typical microwave frequencies around 5-10 GHz, which require operation below 50 mK. 

The large optical pump power not only heats up the cryogenic environment, but also adds noise locally. The presence of a strong pump adds thermal photons to the microwave resonator, which results in added noise in the transduction process~\cite{Xu2021,Sahu2022}. Therefore, the achieavable conversion efficiency is limited by the optical power for quantum operation.

\subsection{Electro-opto-mechanical Transducers}
As a consequence of the aforementioned limitations of the electro-optic coupling, various approaches employ an intermediate mechanical element, leveraging the strong coupling rates between acoustic waves and both optical and electrical radiation. This results in a two-stage conversion process converting first the microwave photons to mechanical phonons with a subsequent transduction to optical photons. This leads to an altered condition for optimum transduction in such a multi-stage process compared to direct electro-optic conversion (cf. above). In order to reach unity conversion efficiency, the electromechanical and optomechanical transduction rates need to approximately match and vastly exceed the intrinsic loss rates of the intermediate mechanical state, i.e. $C_{MO}=\frac{4G_{MO}^2}{\kappa_{M}\kappa_{O}} \gg 1$ and $C_{MMW}=\frac{4G_{MMW}^2}{\kappa_{M}\kappa_{MW}} \gg 1$~\cite{Wu2020}. In practice, phase matching conditions need to be taken into account as well to optimize transduction~\cite{Wang2022} but this is beyond the scope of this paper. Hence, stronger coupling between microwave and optical fields to acoustic waves is possible and also required for efficient overall transduction. Integrated electro-opto-mechanical transducers have shown improved performance compared to integrated electro-optic devices demonstrating microwave-optic entanglement~\cite{Meesala2024}, efficient transduction of  $>$5 \%/$ \mu W$ of optical power sent to the device~\cite{Zhao2024}, and single-shot readout of a superconducting qubit~\cite{vanThiel2025}. These devices can have a small footprint ($<$0.2 mm$^2$) and low average power dissipation, which makes them potentially scalable to hundreds or thousands of transducers on the mixing chamber plate of a dilution refrigerator~\cite{Weaver2024}.

Some of the earliest realizations deployed the radiation pressure force for both electromechanical and optomechanical coupling in a double parametric process via a low frequency (1-15 MHz) mode~\cite{Andrews2014, Arnold2020}. A three-dimensional device using a low-loss membrane mode has still demonstrated the highest overall transduction efficiency~\cite{Brubaker2022}. However, the required microwave pump in these devices with its potential impact on sensitive microwave qubits, the limited bandwidth and difficulty to cool the low-frequency mechanical mode close to its ground state has led the field to look for alternatives. Transducers that use piezoelectric materials such as lithium niobate, aluminum nitride, or gallium phosphide, where an electric field induces mechanical deformation, have recently accomplished strong electromechanical coupling between a wavelength-scale mechanical mode and a microwave resonator, i.e. a coupling rate exceeding both the microwave and mechanical loss rate, without an additional microwave pump~\cite{Jiang2023, Weaver2024, Meesala2024}. These devices are realized by heterogeneous integration of the piezoelectric material and silicon benefitting from the excellent optical and acoustic properties of the latter. The piezoelectric segment and the optomechanical crystal in silicon~\cite{Eichenfield2009} share the same mechanical mode. However, these heterogeneous platforms lead to a more complicated fabrication process and often increased loss rates for microwave, mechanical and optical fields alike. A monolithic electro-opto-mechanical transduction platform with Gigahertz mechanical modes and a large electromechanical cooperativity has been realized by coupling microwave fields to mechanical motions via DC-biased small gap capacitors~\cite{Zhao2024}.

The main limitation for the aforementioned devices originates from the optical pump needed to parametrically enhance the optomechanical coupling rate. This pump induces an excess thermal occupation of the mechanical intermediary mode, which adds noise to the transduction process. Therefore, quantum low noise operation restricts the maximum optical pump powers and thereby the achievable optomechanical transduction efficiency (cf. Table \ref{tab:transducersachieved}). Recent advances in quasi-2D optomechanical cavities demonstrated a significant improvement in pump-related heating resulting in an improvement of the optomechanical conversion efficiency by up to two orders of magnitude  \cite{Sonar2025,Mayor2025,chen2024}. We will discuss the implications of this improvement for future quantum applications in section \ref{subection: projected improvements}.

Finally, other systems can also be used as an intermediary for microwave-to-optics transduction such as magnons and atomic states. The former have strong electro-magnonic coupling, and the latter have strong optical transitions (although typically operating in the lossy visible optical regime). One platform, which was integrated on chip and has achieved recent success is the conversion via the electronic state of an ensemble of rare earth ions with a conversion efficiency of 0.76\% and added noise of 1.24~\cite{Xie2025}.

\begin{table}[ht]
    \centering
    \caption{Examples of two transducers with near-term achievable performance.}
    \begin{tabular}{ccccccp{8mm}p{8mm}}
  Transducer &	$\eta_{MW}$ &	$p_{MO}$ & $\eta_{det}$&$\eta_{tot}$&$	n_{th}$ &$t_{rep}$ $(\mu s)$ & BW (MHz)\\ \hline
 1	&0.8	&0.01&0.5&$4\cdot10^{-3}$&	0.1	& 1 & 10	\\ 
 2	&0.95	&0.1	&0.5	&$5\cdot10^{-2}$&	0.01& 1 & 	10	\\ \hline
    \end{tabular} 
    \label{tab:transducersprojected}
\end{table}

\subsection{Hardware Metrics}
Table \ref{tab:transducersachieved} lists a selection of recent integrated transducer devices and the 3D transducer with the overall best efficiency of $\eta$=47 \%~\cite{Brubaker2022}. The total probability of a microwave photon (from a qubit) being transduced and detected is $\eta_{tot}$. To date, this is far less than unity for all integrated approaches. Nevertheless, especially for the applications discussed in this work, the main figure of merit is the input-referred added noise, $N_{add}$. $N_{add}\lesssim1$ marks the quantum-enabled regime in which entanglement (e.g. qubit-microwave photon entanglement) can be preserved during transduction to create entanglement between a qubit and an optical photon. Thus, $N_{add}$ is a decisive metric combining the conversion efficiency and the thermal occupation or other noise sources of the transducer. As the decoherence times of superconducting qubits are typically on the order of 10s to 100s of microseconds and photon wavepackages from superconducting qubits with high fidelities are on the order of $< 1 \mu s$, the transduction bandwidth (BW) is a key component to process these signals and lies ideally above 1 MHz. A high repetition rate $t_{rep}$ determines the number of transduction processes that can be performed per unit time.

\subsection{Projected Improvements}
\label{subection: projected improvements}

Now we would like to restrict the discussion to two idealized transducer parameter sets that provide the basis to determine the feasibility and performance of optical quantum links between superconducting processors with microwave-to-optics transducers. Transducer 1 in Table \ref{tab:transducersprojected} combines already demonstrated parameters presented in Table \ref{tab:transducersachieved} within one device and is therefore in immediate reach. The only not yet accomplished parameter is non-demolition state transfer between a qubit and a transducer with high efficiency, $\eta_{MW}$. Qubit-to-optical state transfer has only been demonstrated via direct integration, which destroyed the qubit state~\cite{Mirhosseini2020}. However, intermediate steps such as high fidelity state transfer from a qubit to a mechanical resonator~\cite{Satzinger2018,Chu2018} and to a propagating microwave field~\cite{Eichler2012,Kurpiers2019,Ilves2020,Zhong2021,Qiu2024} have already been achieved, as well as non-destructive connection of a microwave-to-optics transducer to the readout resonator of a superconducting qubit \cite{Delaney2022,Arnold2025,vanThiel2025}.

Transducer 2 has a lower thermal occupation and a higher optomechanical conversion rate, each by one order of magnitude. While these metrics are challenging for the current generation of electro-opto-mechanical transducer designs, quasi-2D optomechanical transducers have achieved an optomechanical conversion rate of $>$90\% with a thermal occupation of 0.25 \cite{Sonar2025} or an added noise $<$0.1 with $>$10\% efficiency \cite{Mayor2025,chen2024}. Thus, combining 2D optomechanical devices with the strong electro-mechancial coupling in piezoelectric materials or charge-biased capacitors will present a significant leap towards efficient microwave-to-optics transducers for scalable quantum network applications, and such next generation devices should be within reach shortly. We project that an improved optical detection efficiency, $\eta_{det}$, of 0.5 will likely be possible, motivated by SNSPD detector efficiencies and chip-to-chip coupling improvement~\cite{Alexander2025} and modest improvements in filter insertion losses. We note that the performances of Transducer 1 and Transducer 2 could potentially be achieved by any of the integrated technology platforms described in this section.

\subsection{Summary}

Microwave-to-optics transducers have made substantial progress in the last years. Key foundational capabilities such as the generation of microwave-optics entanglement have been demonstrated~\cite{Sahu2023,Meesala2024,Meesala2024a} and the added noise has been lowered to the extent that quantum experiments are possible. Integrated microwave-to-optics transducers can operate with high bandwidth, which makes them compatible with quantum processor technology such as superconducting qubits. The main limitation in transducer performance remains the efficiency, which is markedly below unity for all integrated approaches in the low-noise regime, but as we will see in the later sections this does not need to prevent them from being used in distributed quantum computing applications.

\begin{figure}
\includegraphics[width=0.95\linewidth]{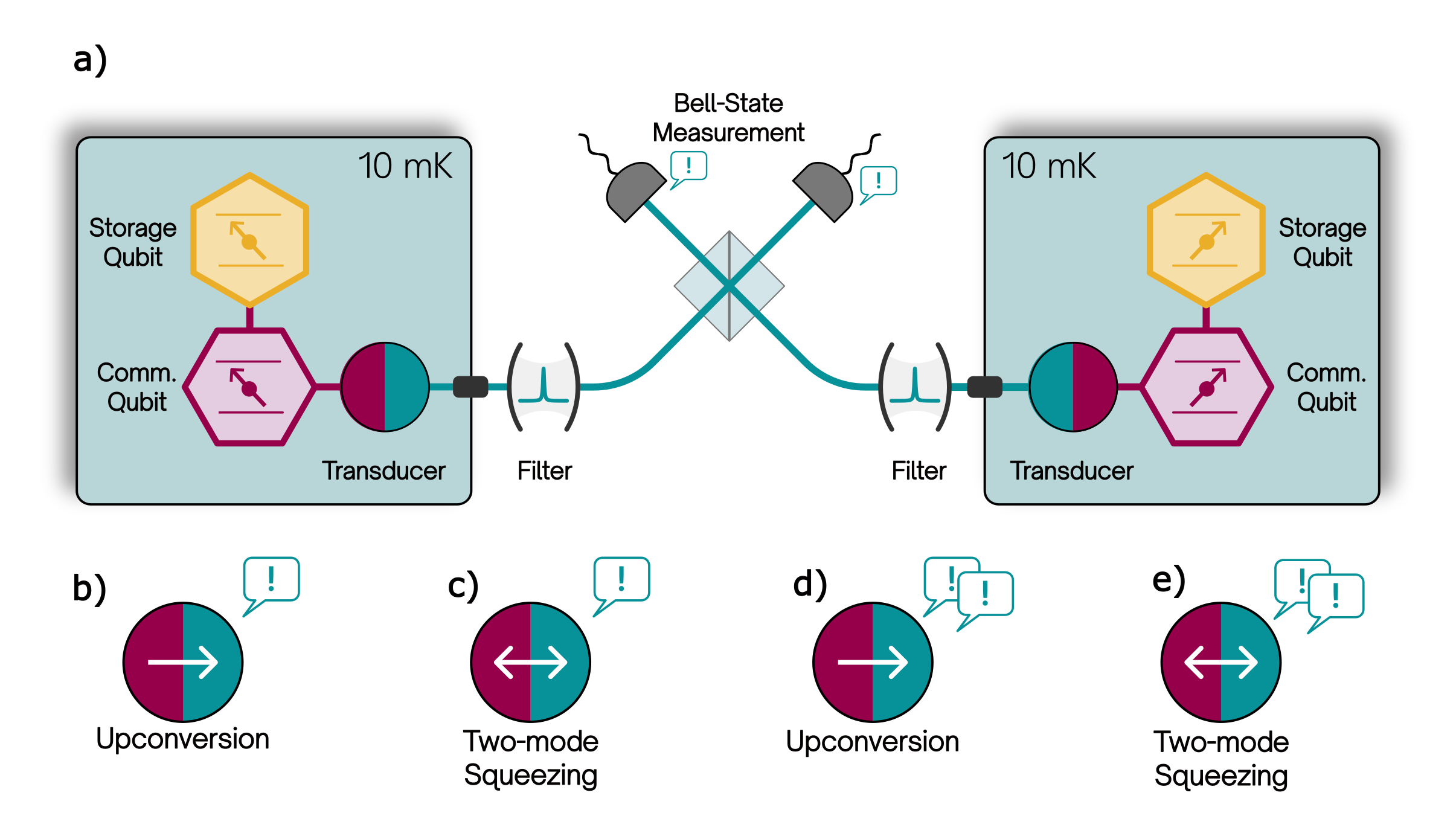}
\caption{Entanglement Generation Protocols. a) General hardware configuration:\ Communication qubits with local qubit storage are linked to microwave-to-optics transducers which in turn are linked to a beamsplitter and two optical detectors, comprising a Bell-state measurement (BSM). The pump for the transducer is generally filtered by an optical cavity. Various protocols for the entanglement generation exist, where we highlight some of the main differences:\ b) One-photon upconversion. A single photon must travel from the communication qubit to the beam splitter from left or right and generate a detection event [!]. c) One-photon two-mode-squeezing (TMS). A single photon must travel from the transducer to the beam-splitter on left or right and a microwave photon is absorbed by a communication qubit. d) Two-photon upconversion.  An early or a late photon must travel from the communication qubit to the beam splitter from the left and the right and generate two detection events [!][!]. e) Two-photon TMS. An early and a late photon must travel from the transducer to the beamsplitter from the left and the right and a microwave photon is absorbed by both communication qubits. Performance of each protocol is discussed in Table~\ref{tab:protocolfidelity}.}
\label{fig:entanglementprotocols}
\end{figure}

\section{Quantum link generation}
\label{section:quantumlinks}

In this section, we investigate the creation of entangled pairs of qubits shared between two or more remote quantum processors in order to perform computing across multiple systems. A number of protocols exist for direct microwave-to-optic-to-microwave transduction from one processor to another or for two microwave-to-optic stages and entanglement swapping via homodyne detection of the optical signals~\cite{Zhong2022}. These protocols come with the downside that they require a total loss rate less than 50\%~\cite{Zhong2022} unless more complicated input states such as squeezed light or GKP states are used~\cite{Wu2021,Shi2024, Wang2024}. Therefore, our main focus will be on a particular set of loss-resilient architectures, to account for loss introduced by the transduction and optical detection inherent to any practical system, as discussed in the previous section. However, many other protocols have been proposed and discussed in literature~\cite{Beukers2024,Davossa2024}. 

Establishing a large number of entanglement links, $N$, between quantum processors will be crucial, because $N$ will scale with the number of qubits per processor and the specific protocol used. If the entangled pairs are delivered with a probability $p_{del}$ to the internode link (see Figure~\ref{fig:overview} blue box upward arrow), then the probability of a successful computation which makes full use of the $N$ links is $p_{del}^N$(Figure~\ref{fig:overview} pink box upward arrow), hence benefiting exponentially from large $p_{del}$. At the same time, realizing a scheme where the entangled pairs can be delivered on-demand, with $p_{del}=1$, despite losses in the conversion chain can significantly help to reach practical quantum link generation, even for large $N$. In addition, on-demand links simplify further abstraction at higher levels of the computing stack, such that quantum computations can be carried out at regular time intervals without reference to individual timing events within each link. For these reasons, we pick two examples using the current and next generation transducers from the previous section and show that these can already achieve on-demand link generation with high fidelity.

\begin{figure*}
	\includegraphics[width = \textwidth,trim={0 12cm 0 1cm},clip]{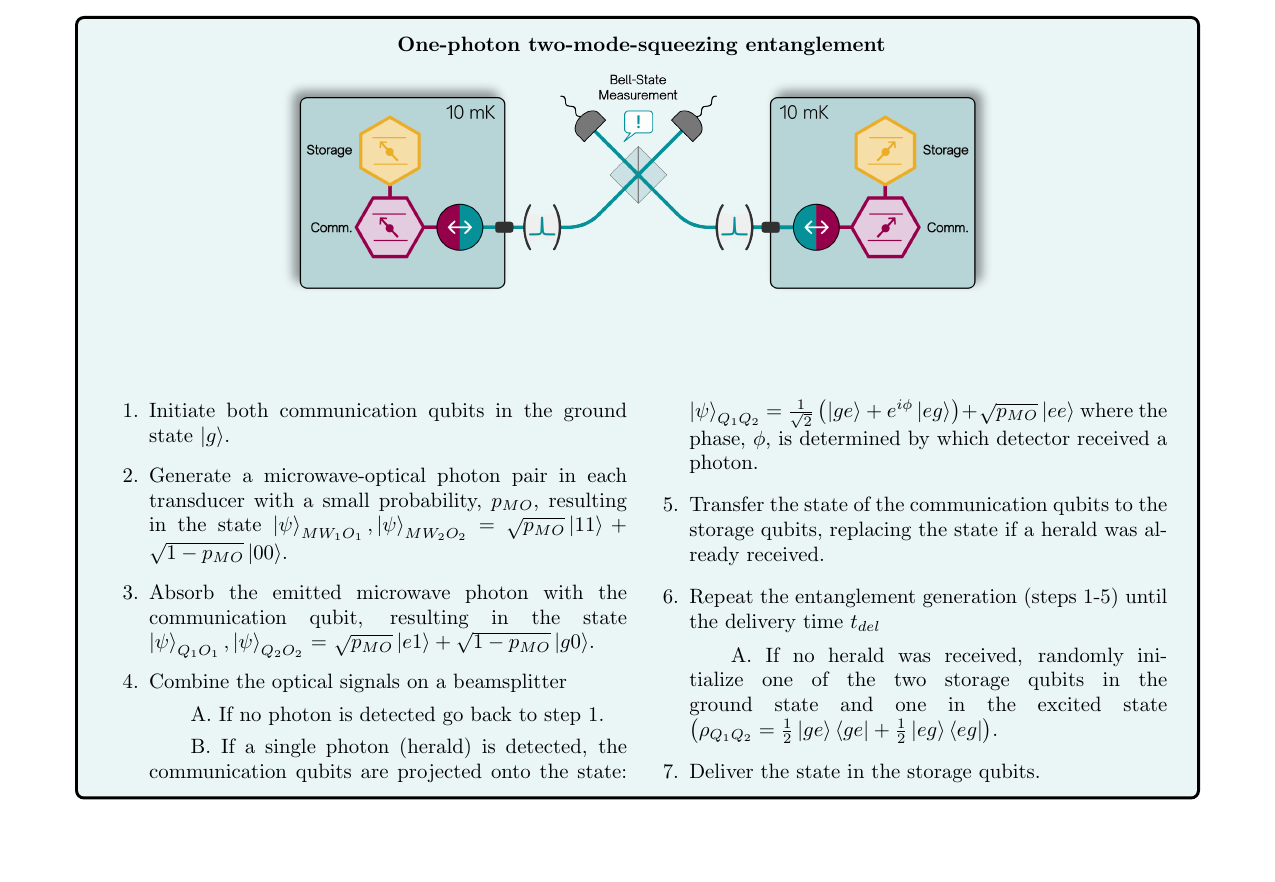}
	\caption{As a first example of how entanglement can be generated, we use a single photon protocol with a two-mode-squeezer (Figure~\ref{fig:entanglementprotocols}c) which is adapted from the DLCZ protocol~\cite{Duan2001}. The protocol is described in detail for the specific case of transducers by Krastanov et al.~\cite{Krastanov2021}.}
	\label{textbox1}
\end{figure*}

\subsection{On-demand entanglement generation} 
\label{subsection:on-demand}

The general format for an entanglement link is shown in Figure~\ref{fig:entanglementprotocols}a. The two communication qubits which are to be entangled are connected via a microwave-to-optics transducer (or optics-to-optics QFC) to an optical beam splitter in the middle. If a strong optical pump is used for the transduction, the pump must be removed with filtering cavities. Local qubit-photon entangled pairs are generated on both sides and the optical photons are sent towards the beam-splitter. A Bell-state measurement (BSM) realised via single photon measurements after the beamsplitter, generates a herald signal, and projects the remote entanglement between the two modules. For the BSM to succeed, the optical photons which are generated by the converters must be indistinguishable in frequency, in their temporal-shape and in polarization. Furthermore, the noise added by the converter and the entanglement protocol must be small, as any added noise will generate false detection events which will herald an incorrect quantum state leading to a low fidelity entangled state.

\begin{figure*}
	\includegraphics[width = \textwidth,trim={0 11cm 0 1cm},clip]{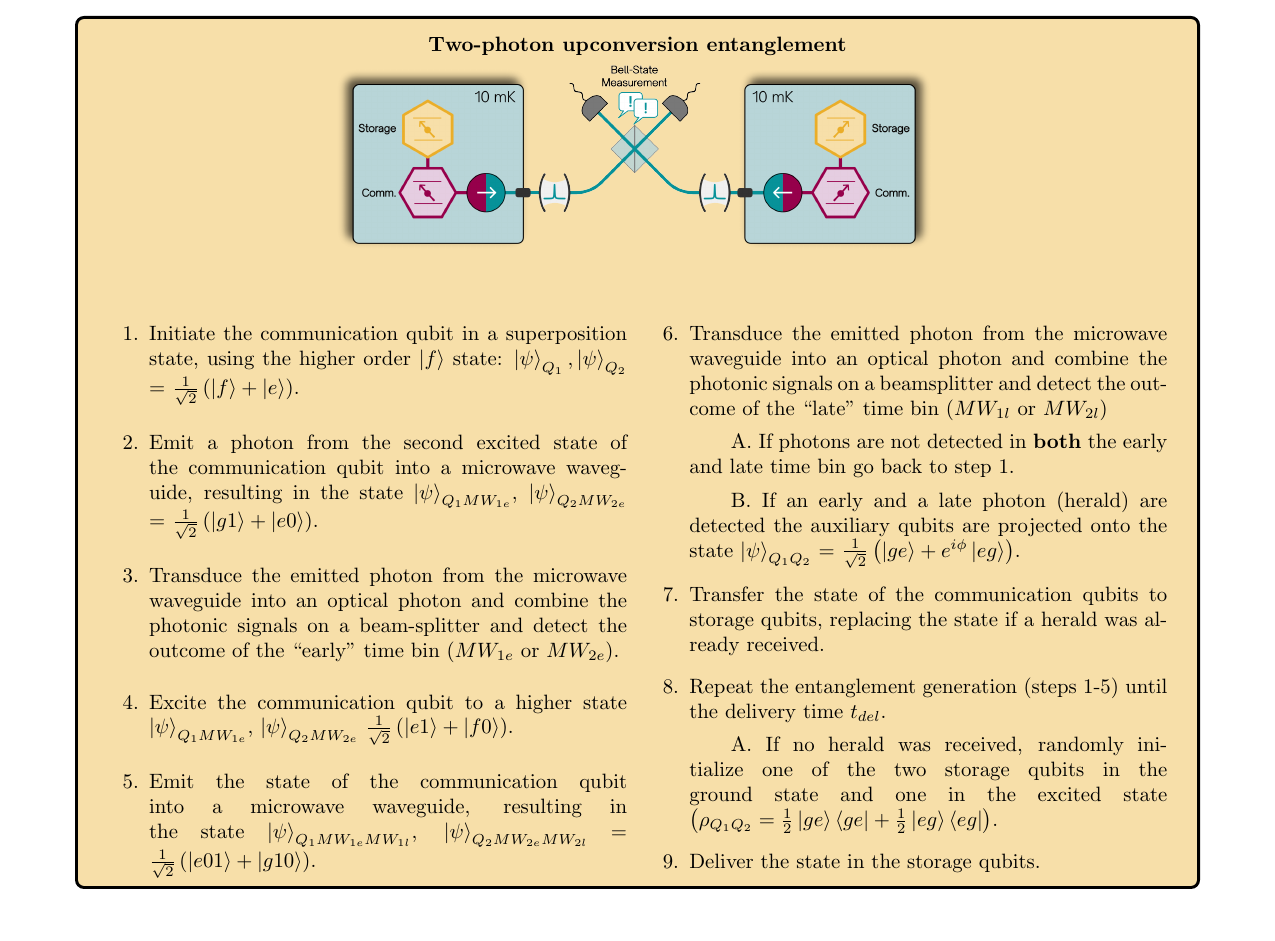}
	\caption{As a second example of how entanglement can be generated we use a two-photon protocol which is adapted from the Barrett and Kok protocol~\cite{Barrett2005}. A variant of this protocol is described in detail for the specific case of transducers by Zeuthen et al.~\cite{Zeuthen2020}. We summarize the key steps of the protocol here. There are multiple ways to emit time-bin photons which are entangled with the state of a superconducting qubit, some of which emit using the readout resonator and second excited state of a transmon~\cite{Eichler2012,Kurpiers2019,Ilves2020} and others which would require an auxiliaray qubit for generating the time-bin state~\cite{Zhong2021,Qiu2024}. All of these methods would in principle work for entanglement generation. We choose to explain one here for clarity.}  
	\label{textbox2}
\end{figure*}

The total efficiency, $\eta_{tot}$, can be split into three distinct parts, which collectivelly determine the probability of heralding a successful entanglement generation, $p_{her}$, within the entanglement link. First, there are losses between the qubit and the transducer and second, there are transduction losses. Finally, there are optical detection losses, which include filtering, optical multiplexing and coupling losses. For most of the transducers discussed in the previous section, these losses are currently still relatively large, leading to $p_{her}\ll1$. In many of these experiments, postselection, a commonly used experimental technique to deal with such loss, is therefore used. Experiments are repeated many times, and the results are conditioned on actual heralding events by discarding unsuccessful attempts. This technique, however, quickly becomes impractical for computations involving a large link number, $N$, as only attempts where all links succeed simultaneously can be used, leading to even smaller probabilities of $p_{her}^N$. 

A protocol which tolerates loss and can support a large number of links is therefore required: real-time feedback with a pre-defined delivery time or timeout, $t_{del}$, after which the entangled state is always delivered ($p_{del}$=1) ~\cite{Hucul2015, Humphreys2018}. In this approach an entanglement attempt is performed and if no herald signal is received from the BSM, the system performs another entanglement attempt after a time $t_{rep}$, which is determined by the repetition rate and bandwidth of the transducer and the qubit's photon emission interface. This process is then repeated for every link independently until a herald signal is received or a timeout is reached. If a herald signal is received the entangled state between the two communication qubits is maintained in a set of storage qubits until the predefined delivery time $t_{del}$\footnote{If no herald is received, the fidelity of the delivered state is maximized by delivering a random classical anti-correlated state which has a 50\% fidelity with the ideal state.}. Therefore, the delivery time or timeout is limited by the storage time of the entangled pair: the decoherence rate of the communication or storage qubits, $T_{coh}\sim T_2$~\footnote{ $T_{coh}= T_2$ if $T_2\ll T_1$}. The link capacity $\eta_{link}=T_{coh} p_{her}/t_{rep}$, should be much larger than 1 to avoid significant reduction in fidelity from decoherence. This has been achieved even in cases with small $p_{her}$~\cite{Humphreys2018} and this type of on-demand entanglement link has recently been used to perform deterministic remote gates between quantum processors as a step in a distributed quantum computation~\cite{Main2025}. Importantly, the storage qubits are always entangled at the end of the protocol, with a fidelity which is reduced by the percentage of repetition blocks in which a herald is not received. Because the qubits are always entangled at the predetermined delivery time, deterministic quantum computations can be run with the quantum link as a resource. The protocol can be applied to multiple links in parallel and entanglement is generated in all $N$ links independently.

\subsection{Entanglement Protocols}

\begin{table*}[ht]
    \centering
    \caption{Heralding probability, protocol infidelity and thermal noise of the 4 protocols. The highlighted blue protocol is outlined in detail in Box~\ref{textbox1}. The highlighted yellow protocol is outlined in detail in Box~\ref{textbox2}. $\eta_{tot}=\eta_{MW}p_{MO}\eta_{det}$ is the total loss in the system from qubit to detector. For the one-photon upconversion scheme the qubit emission probability is $\alpha$. We assume that $\alpha$, $p_{MO}$ and $n_{th} \ll 1$, and therefore omit higher order terms in these parameters.}
    \begin{tabular}{c|cc|cc|cc}
     & \multicolumn{2}{c|}{$p_{her}$} & \multicolumn{2}{c|}{Protocol Infidelity} & \multicolumn{2}{c}{\makecell{Thermal Noise \\ Infidelity}} \\ 
     & 1-photon & 2-photon & 1-photon & 2-photon & 1-photon & 2-photon \\ \hline
    Upconversion & $2\alpha\eta_{tot}$ & \cellcolor{lightyellow} $\eta_{tot}^2/2$ & $\alpha$ & \cellcolor{lightyellow} 0 & $\frac{n_{th}}{\alpha\eta_{MW}}$ & \cellcolor{lightyellow} $\frac{6n_{th}}{\eta_{MW}}$\\ 
    TMS & \cellcolor{lightblue} $\frac{2\eta_{tot}}{\eta_{MW}}$ &  $\eta_{tot}^2/2$ & \cellcolor{lightblue} $\eta_{MW}p_{MO} + (1-\eta_{MW})$ & $ \frac{2}{3}p_{MO}(1-\eta_{MW})$ & \cellcolor{lightblue} $2n_{th}\eta_{MW}^2$ & $2n_{th}$\\ \hline

    \end{tabular} 
    \label{tab:protocolfidelity}
\end{table*}

Here we will briefly focus on 4 specific entanglement protocols, which are similar to the DLCZ protocol~\cite{Duan2001} with the configuration in Figure~\ref{fig:entanglementprotocols}, and discuss their performances. The entanglement protocols are distinguished by two key elements:\ the mode of operation of the transducer and the photon basis. In Fig.~\ref{fig:entanglementprotocols}b-e the four different combinations are highlighted, and a step-by-step procedure is outlined for two combinations in Box~\ref{textbox1} and Box~\ref{textbox2}. 

The first mode of operation for the transducer is as a source of two-mode-squeezing (TMS) entanglement, by pumping with a blue-detuned optical drive as illustrated in Fig.~\ref{fig:entanglementprotocols}c and e. In this case, entangled pairs of microwave and optical photons are generated through a process which is analogous to spontaneous parametric downconversion with a probability of $p_{MO}$~\cite{Aspelmeyer2014}. If the microwave photon is absorbed and its state is imprinted on the communication qubit, then entanglement between the communication qubit and the traveling optical photon is generated. 

The second mode of operation for the transducer is as an upconverter, transferring a quantum state from the microwave to the optical domain by pumping with a red-detuned optical drive as illustrated in Fig.~\ref{fig:entanglementprotocols}b and d. In this case, entanglement is first generated between the communication qubit and a travelling microwave photon~\cite{Eichler2012,Kurpiers2019,Ilves2020,Zhong2021,Qiu2024}. The traveling microwave photon is then sent through the transducer, resulting in qubit-optical photon entanglement. Independent of whether two-mode-squeezing or upconversion is used, entanglement is generated between each remote communication qubit and an optical photon which is sent to a beam-splitter to herald qubit-qubit entanglement.

There are a number of different encodings for flying optical qubits including number state and time-bin qubits. If number states are used (Fig.~\ref{fig:entanglementprotocols}b,c), the optical state is encoded in the number of photons and entanglement is heralded between the two communication qubits if one of the detectors after the beamsplitter detects a single photon. As only a single detection event is required, the generation of an optical photon from one of the two qubits only needs to succeed once. However, if microwave-optic entanglement is generated on both sides in the same attempt, the procedure fails, decreasing the fidelity.

An alternative is to use time-bin encoded qubits. The entanglement protocol then involves two qubit-optical photon generation steps separated by a delay time (Fig.~\ref{fig:entanglementprotocols}d,e). The protocol prepares the optical state as a photon in a superposition of the early or the late time-bin, which is entangled with the state of the qubit. Entanglement is heralded between the two quantum processors if both an early and a late photon are detected after the beam splitter, making this encoding more resilient to noise \footnote{In the two-mode squeezing case, two qubits are necessary in each link which store early and late component and then perform a heralding parity measurement to ensure that early and late photons were not generated by the same transducer \cite{Zhong2022}.}.

Table~\ref{tab:protocolfidelity} shows a comparison of the protocols and their heralding probabilities and infidelities. One-photon protocols have a significantly higher heralding probability as the losses of the transduction chain are only included once. However, these protocols have a higher protocol infidelity from the probability of exciting both qubits. TMS schemes suffer a decrease in fidelity from any microwave losses between the transducer and the qubit ($\eta_{MW}$), but have less susceptibility to thermal noise, $n_{th}$, because the quantum signal originates at the same location as the noise as opposed to the upconversion case where the quantum signal is attenuated before the noise is added. Depending on the particular parameters of the transducer and the communication qubit and the required link fidelity, different protocols may be optimal for the entanglement generation.

\subsection{Boosting rate with quantum repeaters}

\begin{table*}[ht]
    \centering
    \caption{Heralding probability increase with optical memories.}
    \begin{tabular}{cccc}
     Protocol & $p_{her}$ no memory &  $p_{her}$ with memory  & Probability Increase \\ \hline
     \rowcolor{lightyellow} 2-photon upconversion, spin-cavity & $(\eta_{tot}^2)/2$ & $(\eta_{tot}\eta_{mem})/2$ & $\eta_{mem}/\eta_{tot}$ \\ 
    2-photon TMS, catch and release & $(\eta_{tot}^2)/2$ & $(\eta_{tot}\eta_{MW}\eta_{mem}^2)/2$ &$\eta_{mem}^2\eta_{MW}/\eta_{tot}$\\ \hline 
    \end{tabular} 
    \label{tab:memenhancement}
\end{table*}

A quantum repeater is an element which receives a quantum signal and re-outputs that same quantum signal at a controllable time. This can be used to extend the reach of long distance quantum networks, but also can be useful to boost entanglement rates in a local network with losses~\cite{Briegel1998,Azuma2023,Wehner2018}. Scalable microwave-to-optics converters and their associated filtering and detection hardware will likely always have some inefficiencies. Quantum repeaters boost the heralding rate and reduce the impact of loss by removing the need for simultaneous success when operating two-photon detection protocols. Here we will investigate several generic configurations of a quantum repeater which can be used to boost the rate of entanglement generation between quantum processors. Each memory has an efficiency of storing and then re-emitting a photon, $\eta_{mem}$, which must be close to 1 to effectively boost the signal. While some quantum memories operate at telecommunications wavelengths~\cite{Askarani2019,Rakonjac2022, Thomas2024, Wang2025, Wallucks2020}, the majority of quantum repeaters to date have operated at visible wavelengths. As such the memory efficiency may include an extra frequency conversion step for matching the memory wavelength to the microwave-to-optics transducer wavelength~\cite{Krutyanskiy2023,Tchebotareva2019,vanLeent2020}. A number of physical platforms have been able to achieve high memory efficiency~\cite{Azuma2023,Bhaskar2020,vanLeent2020,Ma2022,Wang2025, Vernaz-gris2018}. 

In particular, we will discuss three configurations for a quantum repeater. Firstly, a storage qubit can be used as a microwave quantum memory. The two qubit register (communication qubit and storage qubit) used in section~\ref{subsection:on-demand} to boost the heralding probability through repeated entanglement attempts can therefore be the basis for a quantum repeater. A recent work explored how the quantum memory can be used to enable other protocols with favorable scaling, such as the extreme photon loss protocol~\cite{Dirnegger2025,Campbell2008}. The storage qubit can also be replaced by longer lived microwave or phononic modes. In principle, one can create two entanglement links between a middle quantum network node and two neighboring quantum processors and then combine them in the middle node to generate a longer entanglement link: a microwave-memory based quantum repeater. However, further transducer improvements in efficiency are required to be competitive with optical-memory based quantum repeaters for long distance networks.

\begin{figure}
\includegraphics[width=0.95\linewidth]{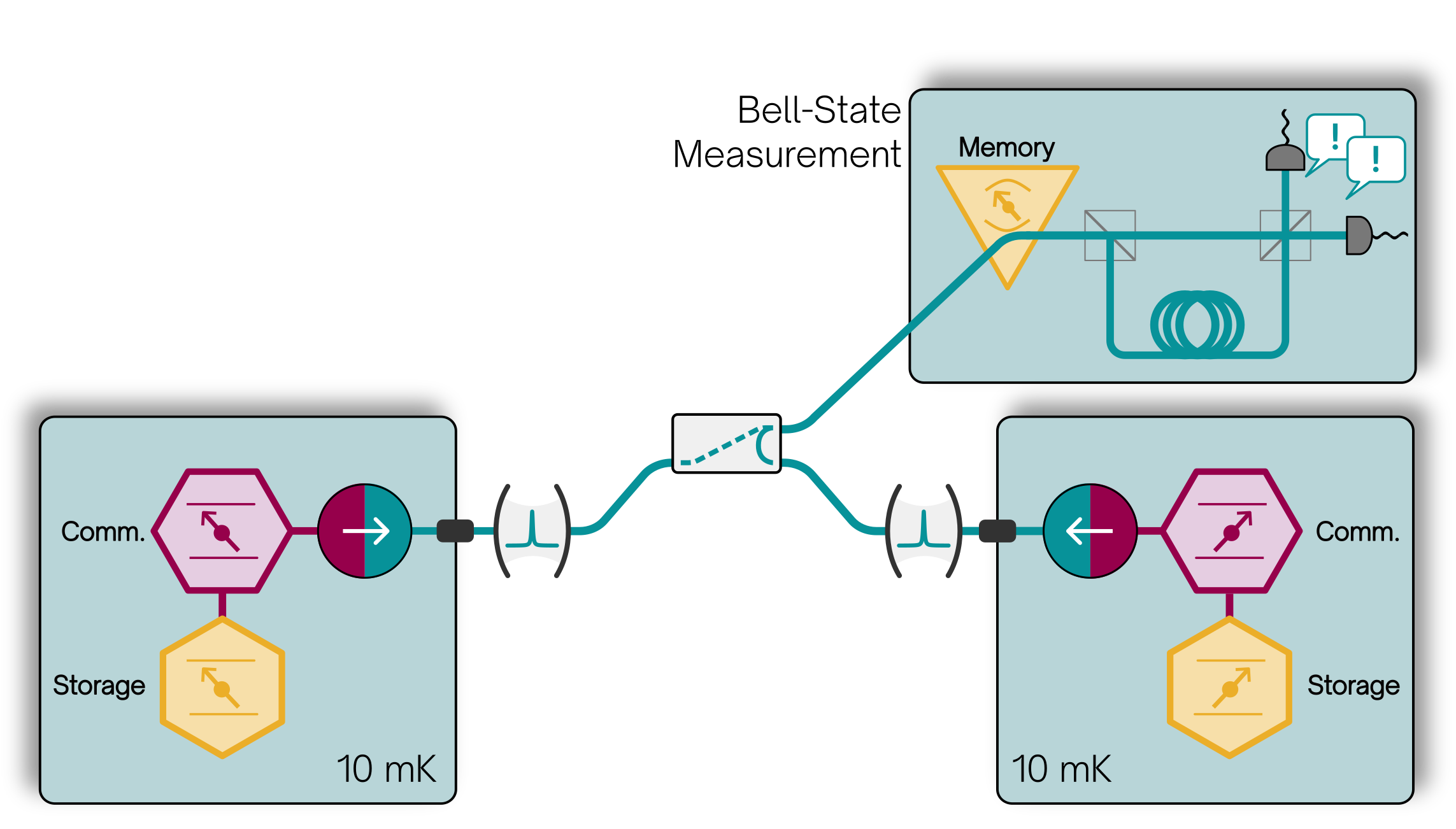}
\caption{Boosting entanglement rates with optical memories. In this figure we illustrate a configuration in which the two-photon upconversion scheme can be expedited. The beam splitter is replaced with a fast optical switch and a spin-cavity optical memory and a time-bin interferometer which erases the time of arrival of the heralding photons.}
\label{fig:memoryprotocols}
\end{figure}

In addition, entanglement protocols can be enhanced by including optical memories. One type of optical memory is a qubit-cavity memory~\cite{Duan2004,Reiserer2014, Nguyen2019}. The cavity is used to imprint the state of a traveling photon onto a stationary qubit, which subsequently stores the quantum information. If an appropriate qubit basis is used, the traveling photon can be detected following the interaction to herald that a storage event took place. This geometry has previously been used to increase the optical channel capacity~\cite{Bhaskar2020}. The same experimental configuration can replace the BSM in the previous section and result in an enhancement of the entanglement rate if a two-photon protocol is used. For example: the memory can be switched back and forth between the two (filtered) transducer outputs until a herald is received from both transducers (see Figure~\ref{fig:memoryprotocols}). The combination of two heralds and a local qubit measurement projects the remote qubits into an entangled state with the added advantage that the photons from the two interferometer arms no longer need to arrive simultaneously~\cite{Bhaskar2020}. As we described in the “on-demand” section, the chosen timeout (or delivery time) $t_{del}$ should be shorter than both the memory lifetime and the storage qubit lifetime. Coherence times of 1 ms have been achieved by memories in this geometry~\cite{Nguyen2019}.

A second type of optical memory is a catch and release memory. Catch and release memories have been built out of a number of physical systems, including rare-earth ion doped crystals and clouds of trapped atoms. In clouds of trapped atoms, a control pulse can “write” or “read” a single collective atomic excitation within the cloud~\cite{Cao2020}. However, these memories do not have an optical herald, so they need to be combined with a herald to indicate succesful transfer in one arm of the interferometer. Replacing a beam-splitter with catch and release memories can enhance the two-photon TMS scheme discussed in the previous section, where the transducers are used as a source of entanglement. A requirement of this scheme is that both quantum processors herald the acceptance of an appropriate state by performing a parity check on the received microwave qubits as has been proposed by Zhong et al.~\cite{Zhong2022}. Because the two transducers do not need to succeed simultaneously, the two-photon TMS protocol is sped up by $\eta_{tot}$ and slowed down by the microwave heralding efficiency $\eta_{MW}$ and the efficiency of the memories on both sides $\eta_{mem}$. Table~\ref{tab:memenhancement} summarizes the performance advantages of optical memories. As is the case for spin-cavity memories, the delivery time, $t_{del}$ should be shorter than both the memory lifetime and the storage qubit lifetime.

\begin{table*}[ht]
    \centering
    \caption{Two sets of possible transducer and memory performances which are within reach of current technologies, where the second set includes modest improvements in transducer performance to existing devices.}
    \begin{tabular}{c|cccccc|cc}
& \multicolumn{6}{c| }{Transducer} & \multicolumn{2}{c}{Microwave Memory} \\ 
Transducer &	$\eta_{MW}$ &	$p_{MO}$ & $\eta_{det}$&$\eta_{tot}$&$	n_{th}$ &$t_{rep} (\mu s)$&$T_2 (\mu s)$&$T_1(\mu s)$\\ \hline
1	&0.8	&0.01&	0.5&	4$\times10^{-3}$&	0.1	& 1	&200 & 500\\ 
2	&0.95	&0.1	&0.5	&5$\times10^{-2}$&	0.01& 1	&2500 & $10^5$\\ \hline
    \end{tabular} 
    \label{tab:entanglementparameters}
\end{table*}

\subsection{Boosting rate with parallelization }

Another method to increase the speed of the link is to use multiple transducers in parallel as shown in Figure~\ref{fig:parallelizationdistillation} top. In the simplest case, the two storage qubits from Figure~\ref{fig:entanglementprotocols} are supplemented by 2$N$ communication qubits and 2$N$ microwave-to-optics transducers. The $N$ communication qubits in each quantum processor are each connected to a microwave-to-optics transducer and directly or indirectly to the shared storage qubit. Each pair of communication qubits attempts to generate entanglement and transfers its state to the storage qubit if successful. The storage qubit then delivers the state at the end of the delivery time. This technique increases the herald probability by a factor of $N$ (for small $p_{her}N$), but requires more qubits in the quantum processor to be dedicated to interprocessor communication. Technologically, parallelization favors multimode optical memories and transducers which are densely packed and have limited local heat dissipation.

\subsection{Boosting fidelity with entanglement distillation} 

\begin{figure}
\includegraphics[width=0.8\linewidth]{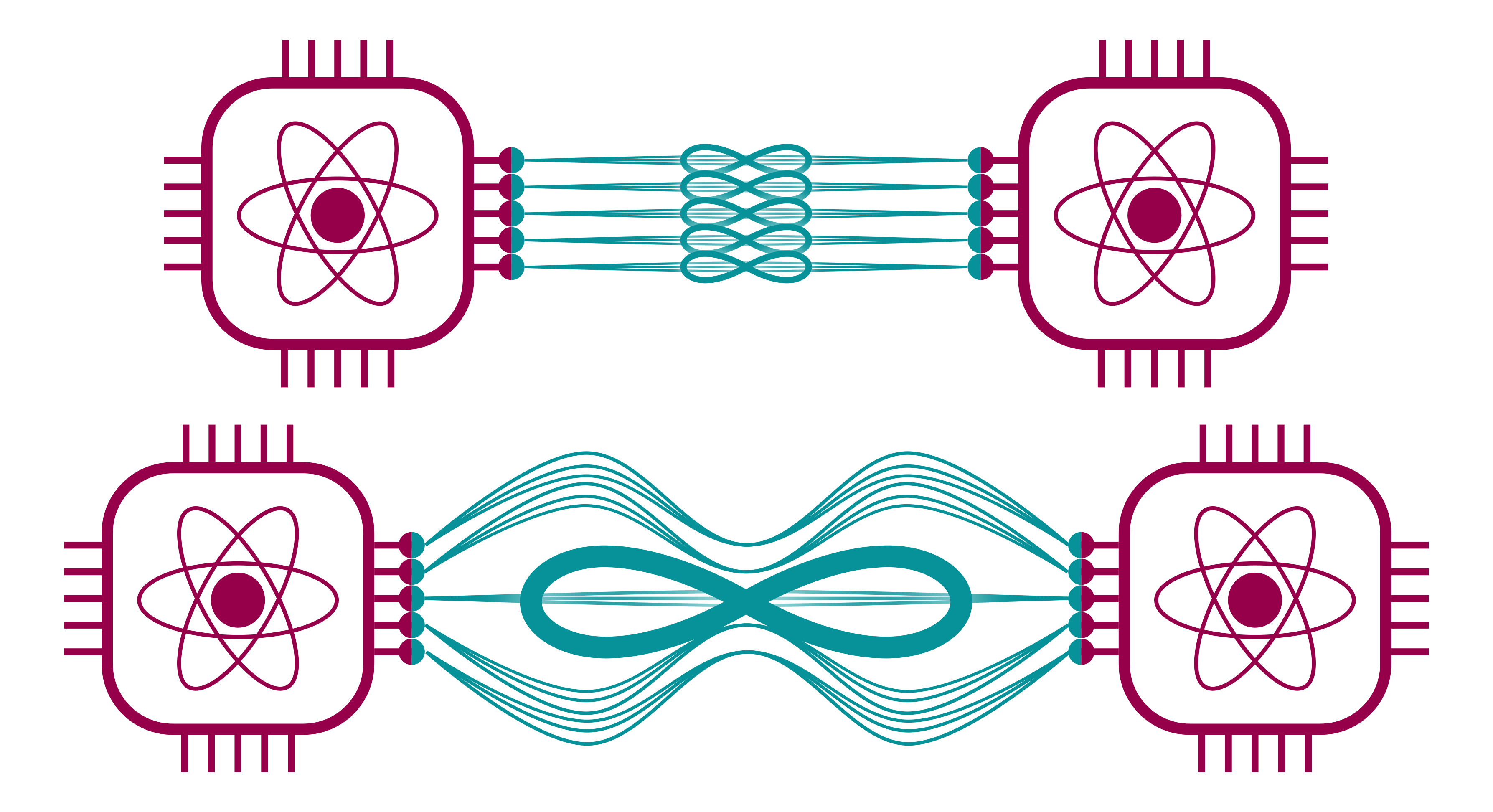}
\caption{Through parallelization (top) multiple entanglement links are generated in parallel using a number of qubits in each quantum processor. This results in a higher heralding rate of a smaller number of links between the processors. Using distillation (bottom) multiple entanglement links are generated in parallel using a number of qubits in each processor. Once entanglement has been created, the individual remotely entangled qubits are locally combined to generate a smaller number of links with higher fidelity.}
\label{fig:parallelizationdistillation}
\end{figure}

In some cases, the link may require a higher fidelity than is possible in a single entanglement generation attempt. If this is the case, entangled pairs with low fidelity can be consumed via local operations in each processor to distill out a single high fidelity entangled pair between the two remote processors as shown in Figure~\ref{fig:parallelizationdistillation} bottom. This process of entanglement distillation has been demonstrated experimentally~\cite{Kalb2017} and it has been shown theoretically that the infidelity of the entangled pairs can be reduced by orders of magnitude, if enough input pairs are provided~\cite{Ang2024}. 

The process of entanglement distillation requires more hardware resources because additional transducers and communication qubits are required for each input to the distillation process. Nevertheless, if processors have enough qubits dedicated to interprocessor communication and if transducers can be operated in large quantities, higher fidelity (on-demand) pairs can be delivered. This can be essential for more advanced protocols such as fault-tolerant surface code expansion as we will discuss in the next section.

\subsection{Example link configurations with transducers}
\label{subsection:examplelinks}
 
Now that we have outlined protocols for entanglement links, we can evaluate the performance with three example configurations based upon the performance of the microwave-to-optics transducer from Section~\ref{section:transducerreview}. We assume a repetition rate of 1 MHz (set by the transducer bandwidth, and the communication qubit reset time). With modest improvements to integrated transducers, a total transduction efficiency of 0.01, thermal noise of 0.1 and a microwave loading efficiency of 0.8 is well within current experimental reach (Table~\ref{tab:entanglementparameters} transducer 1). 

\begin{figure}
\includegraphics[width=0.95\linewidth]{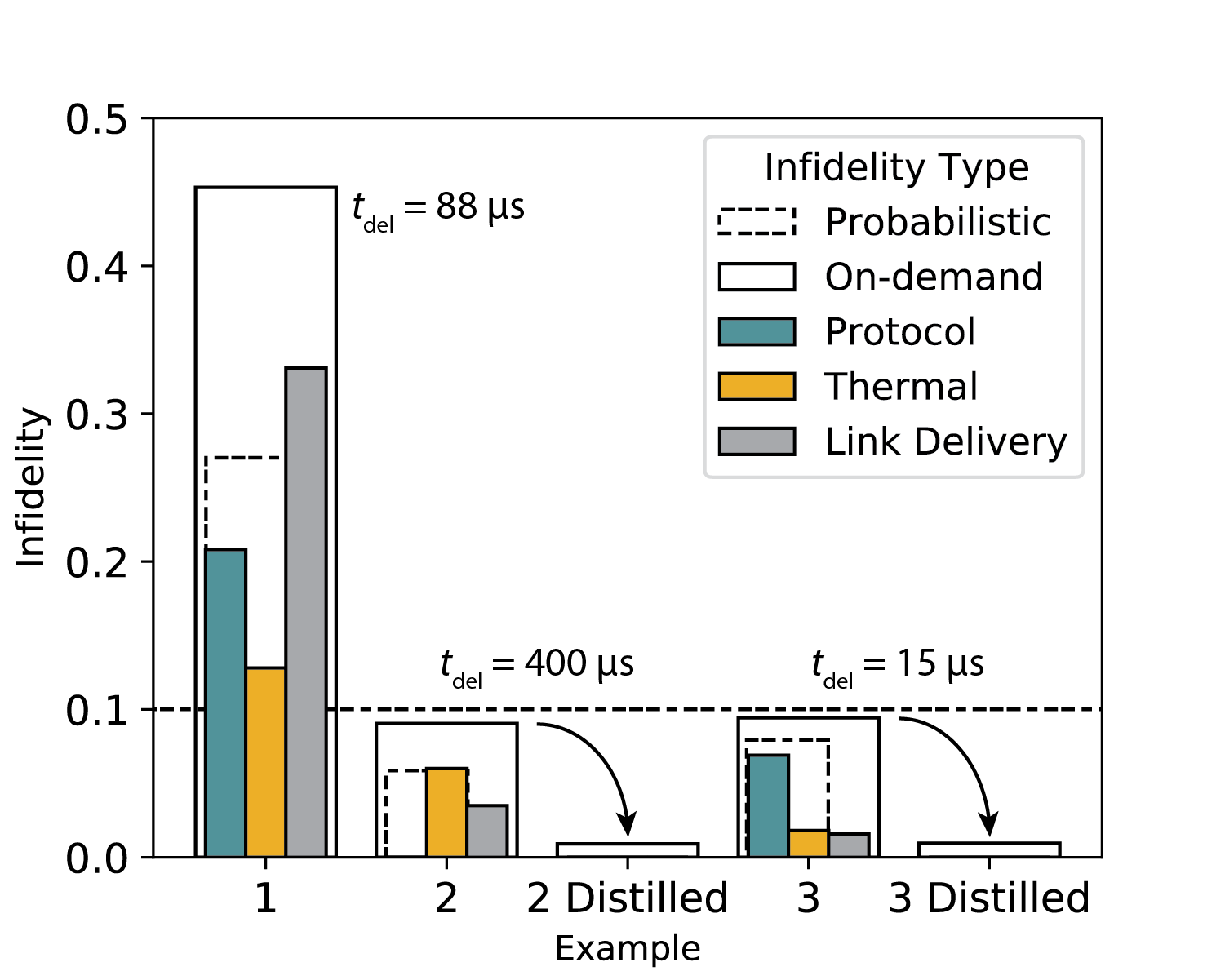}
\caption{Infidelities for the different schemes. We break down the contributions to infidelity for probabilistic entanglement (delivery without a timeout and immediately after a herald) and on-demand at the specified $t_{del}$ in the plot for each example in Table~\ref{tab:entanglementselections}. The dashed line at 0.1 gives the threshold for surface code error correction with lattice surgery (see Section~\ref{section:architecture}).}
\label{fig:infidelitybarplot}
\end{figure}

In Example 1, we explore a link with transducer 1 using the one-photon, TMS protocol. The entanglement link has a protocol infidelity of 0.21 and a thermal infidelity of 0.13 which results in a fidelity of 0.73 for states which are heralded with the link. On-demand entanglement with fidelity above the classical threshold (0.55) can be delivered in a time of 88~$\mu$s (Table~\ref{tab:entanglementselections}). The various infidelity contributions and delivery times for each example are presented in Figure~\ref{fig:infidelitybarplot}. Therefore, transducers with these parameters are already sufficient to show the process of entanglement generation, while using the links will likely be probabilistic.

\begin{table*}[ht]
    \centering
    \caption{Examples of several protocols which could be chosen for each parameter set. *Note that in Example 3, we intentionally reduce $p_{MO}$ to 0.02 in order to reduce the protocol infidelity. Examples 2 and 3 can be boosted by 4 rounds of distillation which increases $N$ by a factor of 16.}
    \begin{tabular}{ccccc}

Example& Parameter Set &	Protocol &	$N$ &$t_{del} (\mu s)$\\ \hline
\rowcolor{lightblue} 1	& 1& 1-photon TMS	&	1	& 88\\ 
\rowcolor{lightyellow} 2	& 2& 2-photon upconversion with memory	&	1 ($\times$16)	&400\\ 
\rowcolor{lightblue} 3	& 2* & 1-photon TMS with parallelization & 20 ($\times$16) & 15 \\ \hline
    \end{tabular} 
    \label{tab:entanglementselections}
\end{table*}

We project that, with already demonstrated improvements to thermal anchoring, transducer 2 will achieve a transduction efficiency of 0.1, thermal noise of 0.01 and a microwave loading efficiency of 0.95 (Table~\ref{tab:entanglementparameters}). In Example 2, we explore the possibility of combining transducer 2 with a qubit-cavity optical quantum memory and a longer lived microwave storage mode with a coherence time of 2.5~ms, which is achievable using qubits or mechanical memories~\cite{Somoroff2023,Bland2025,Milul2023, Liu2024}. Due to the presence of the memories, the two-photon red detuned protocol can be used with higher efficiency, with a heralding probability of $3\times10^{-2}$ and a delivery time of 400~$\mu$s. Here, the entanglement fidelity is limited by thermal infidelity, link efficiency and decoherence to 91\%. The thermal noise from the transducer is the largest contribution.

In Example 3, we combine 20 parallel links with transducer 2 and the microwave storage with 200 $\mu s$ coherence time. Because the transducer scattering rate can be very high and this limits the protocol fidelity of one-photon TMS, we intentionally lower $p_{MO}$ to 0.02, which reduces $\eta_{tot}$ to $10^{-2}$. The initial fidelity of each link is 91\% in a delivery time of 15 $\mu$s, limited primarily by the protocol infidelity. 

By consuming 16 Bell pairs for four rounds of entanglement distillation, the infidelity can be reduced by an order of magnitude (for both Examples 2 and 3), resulting in a net fidelity of 99\%~\cite{Ang2024}. For Example 2, this requires 16 pairs of transducers and 16 memories, and for Example 3 this requires 320 pairs of transducers and 320 memories in each refrigerator. If the distillation operations are performed during the next link generation step, distillation will only increase the delivery time by the two-qubit gate time, which is negligible. This further demonstrates the simplicity of using on-demand entangled pairs for computations.

\subsection{Summary}
\begin{figure}
\includegraphics[width=0.95\linewidth]{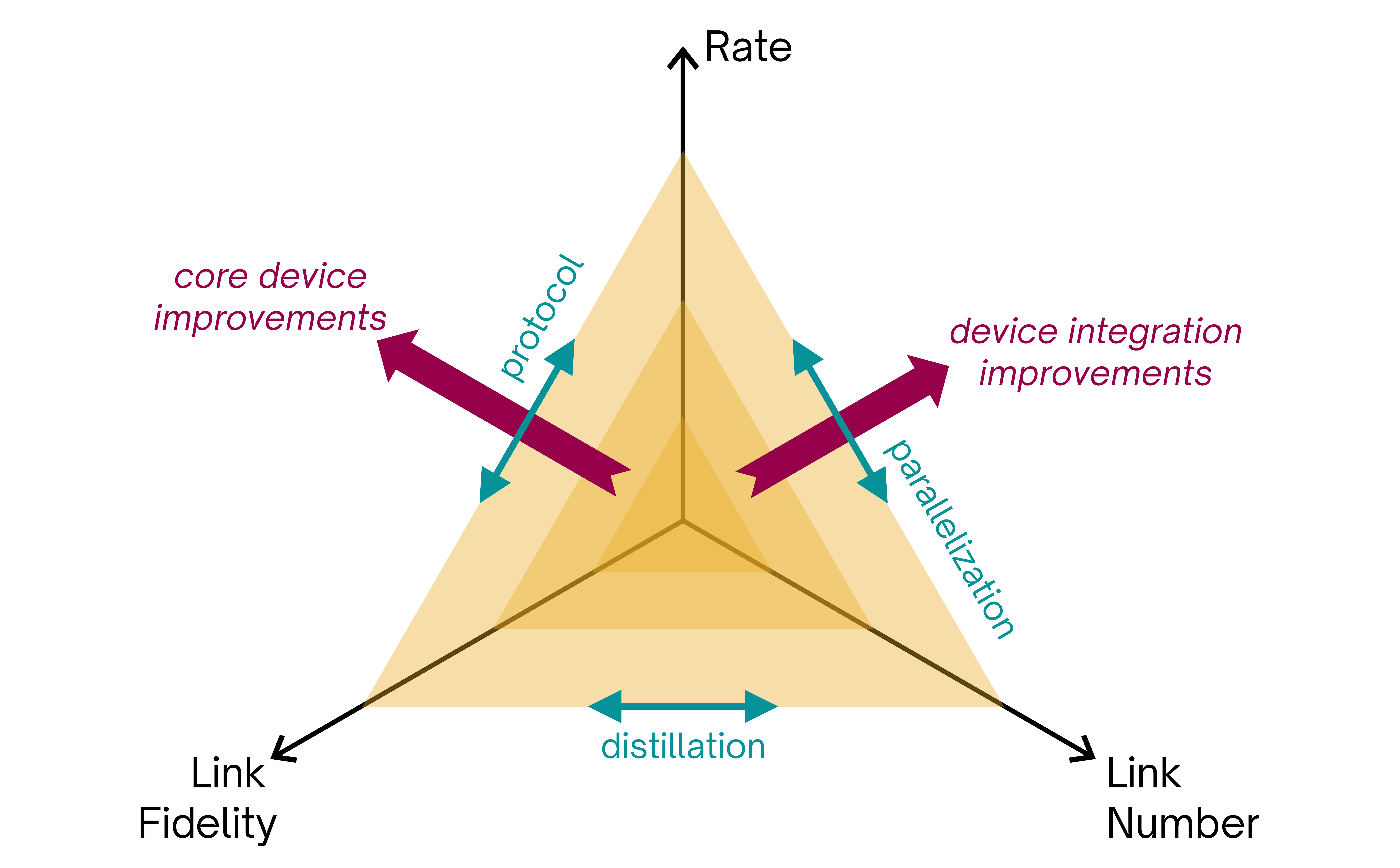}
\caption{In order to achieve a target link number, link rate and link fidelity, a large number of transducers operating quickly and with high fidelity will be required. It is likely that one of the three will limit the achievable target protocol performance. This constraint can be alleviated such that the full capacity of the system is used with three trade-offs. Link fidelity and link rate can be interchanged by the choice of protocols and execution parameters. Link rate and link number can be interchanged by using more or less transducers per link via parallelization. Finally, link number and link fidelity can be exchanged by consuming links to enhance the fidelity through entanglement distillation.}
\label{fig:entanglementtradeoffs}
\end{figure}

While microwave-to-optics conversion with simultaneous high bandwidth, close to unit efficiency and integration still requires significant performance improvements, we demonstrate how it is possible to boost entanglement delivery rates beyond the rate of decoherence for on-demand delivery of entangled states with only modest qubit memories, which are available today. A number of different entanglement schemes can be used, depending on the transduction parameters and the desired fidelity or rate. Further improvements, such as parallelization, distillation and additional optical memories,  can be used to boost the fidelity or rate, and also give further flexibility to trade off between fidelity, number of channels and rate (see Figure~\ref{fig:entanglementtradeoffs}). Given these scenarios and existing transduction technologies, on-demand entanglement between superconducting quantum processors is a realistic outlook and given current and future developments which are taking place with microwave-to-optics transducer technology, scaling of quantum computing across multiple processors is well within reach.

\section{Architectures: late and early stage}
\label{section:architecture}

Quantum computers will expand their computing power greatly if they can incorporate multiple quantum processors. As we explored in the previous section, this can be done in a self-contained and predictable manner with quantum transducers and other supporting hardware as the backbone for quantum links. High fidelity, fast and plentiful links between processors are highly desirable, but the exact requirements depend on the architecture of the quantum computer. In general, as long as the links bring added functionality and/or algorithmic scaling to the quantum computer which cannot be achieved by classical combination methods such as circuit cutting, it will be worth it to incorporate the links. In this section we summarize proposed architectures which will achieve a benefit to scaling and determine if they are feasible in the short term with present transducers.

\subsection{Lattice surgery}
\begin{figure*}
\includegraphics[width=0.8\linewidth]{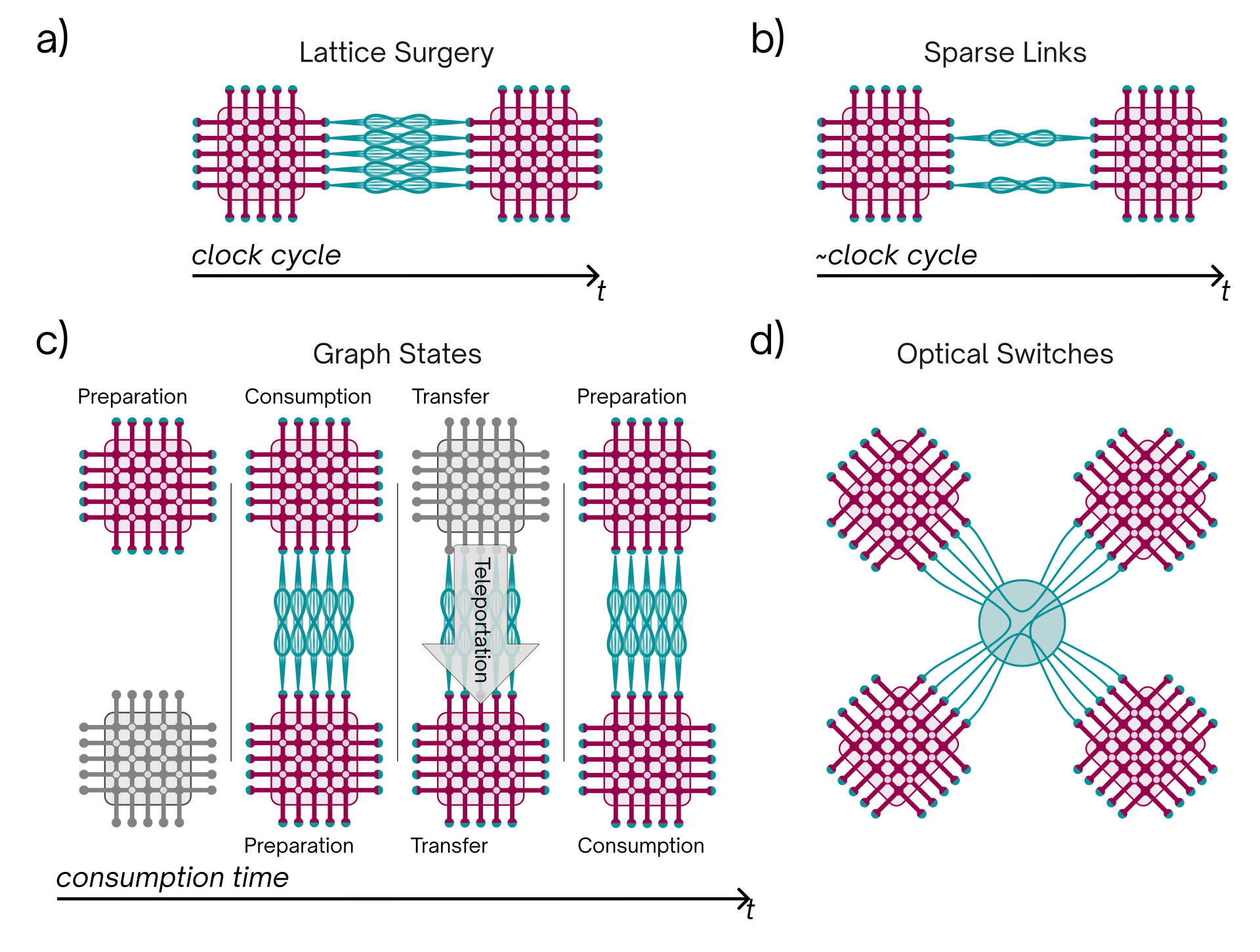}
\caption{Four potential architectures. a) In lattice surgery all edge physical qubits are connected to each other at every clock cycle. b) Sparse links (in space and time) are less resource intensive and could still provide scaling benefits. c) Graph state generation and consumption uses all edge physical qubits, but only requires link generation at fewer times. d) Optical switching enables dynamical reconfiguration during a computation.}
\label{fig:architectures}
\end{figure*}

Most quantum computing companies are pursuing 2D error correcting codes such as the surface code. These codes have recently demonstrated that they are able to reduce the error in logical qubits below that of the constituent physical qubits~\cite{Acharya2025}. One approach to extend these codes is to simply “knit” together two patches from different quantum processors, which is also referred to as \textit{lattice surgery}. Combining these two patches requires quantum links between individual physical qubits at the edge of each quantum processor (see Figure~\ref{fig:architectures}a). We will discuss in this section the requirement for microwave-to-optics transducers to generate these links.

Lattice surgery with full functionality requires that all physical qubits are connected at every clock cycle of the error-correction cycle. This allows for error correction of logical qubits which are shared between the remote processors, correcting for both “space-like” and “time-like” errors. In general making links at the clock cycle of the processor is a far more demanding goal than “on-demand” entanglement, because the clock cycle is typically much shorter than the lifetime of physical qubits. Fortunately, the qubit links have a $\sim$10 times lower error threshold for fault-tolerance than that of the bulk qubits~\cite{Ramette2024}. Thus the memory-enhanced examples from the previous section already have the fidelity required to surpass the error correction threshold (90\%) and to safely exceed it (99\%) with just a few rounds of entanglement distillation.

To achieve clock-cycle delivery (with $t_{del} = 1 \mu$s) in the examples from the previous section, will require a 15-400 times increase in link speed. In principle, this can be achieved with parallelization with 15-400 times as many qubits and transducers (and memories) per link. In the most favorable case where no entanglement distillation is used, the clock-cycle links would use 300 or 400 qubits and transducers per processor per link for Example 2 or 3 respectively (transducer 2, Table~\ref{tab:entanglementselections}). For a quantum processor with 1,000 qubits, one would need to connect approximately 32 edge physical qubits. With some of the transducer geometries presented in Section \ref{section:transducerreview}, it is in fact possible to reach more than 10,000 transducers per module. However, additionally, more than 10,000 qubits would be necessary to generate the links, exceeding the size of the entire processor itself.

Therefore, with near-term performance levels, transducers will not directly be able to support full lattice surgery until further improvements in transduction efficiency and repetition rate are implemented. The product of the total system efficiency ($\eta_{tot}$) and the repetition rate must increase by a factor of 10-100 (see Section~\ref{section:transducerreview}). Fortunately, the bandwidth of a number of transducers already exceeds the 1 MHz repetition rate, which we considered in our examples, by as much as an order of magnitude~\cite{Warner2025,Weaver2024}, allowing for a clear path to optical links even for full lattice surgery, once additional added noise at such high rates can be mitigated.

The lattice surgery approach is a particularly unfavourable way of connecting quantum processors together if the links are the weakest element in terms of rate and fidelity, as the links are put on an equal footing with local two-qubit gates. The next sections will explore other configurations that lower these requirements on internode links with respect to local clock-cycles.

\subsection{Sparse link generation}
Fortunately, it is not necessary to fully connect the edges of error-corrected remote quantum processors to significantly extend the computational capabilities. Sparse remote connections can be used to increase the quantum volume of a computer or to transform a two dimensional error correction code into a higher dimensional, less resource intensive code (see Figure~\ref{fig:architectures}b).

Connections between remote superconducting processors via optical links have been explored by Ang et al.~\cite{Ang2024}. They performed simulations which explored the collective advantage of using two processors together. Indeed, if an optical link with sufficient fidelity and generation rate could be produced, they found that the quantum volume and algorithmic performance with ADDER, Bernstein-Vazirani, quantum Fourier transform and GHZ state distribution could be improved in a system with two quantum processors composed of 5 qubits each. In the previous section we estimated that near-term transducers could deliver links with fidelities above 90\% “on-demand” in a period of 15-400 $\mu$s. This falls within the region where there is an advantage for GHZ state generation and the Bernstein-Vazirani benchmark, but would not yet improve the other metrics or quantum volume. In principle improving the distilled fidelity from 99\% to 99.9\% or increasing the delivery rate for 99\% fidelity by a factor of 10 would be sufficient to increase the quantum volume in the 5 qubit processor considered above. However, it is likely that quantum processors with at least 1,000 qubits will soon be available. Therefore, an immediately interesting yet still open question is of how these links will perform on bigger processors and if additional advantages can be achieved. While it is challenging to simulate this directly with available classical resources, on such a processor it will be possible to allocate fewer links per qubit, which will reduce the overall communication overhead. Multiple links in parallel will also bring additional advantages.

One approach to connect remote circuits is to repeatedly run computations on both sides to either simulate remote gates or mitigate errors in a quantum link. Ang et al. also consider the scaling conditions for which a quantum link which is enhanced by error mitigation outperforms a classical link which is generated with circuit cutting techniques. They use the number of circuits required to simulate a perfect link, which scales as $\gamma^k$, where $k$ is the number of links as a metric. They found that quantum link generation had a better scaling coefficient, $\gamma$, as long as the infidelity of the generated states was below 30\%. With an infidelity of 10\%, it will therefore be possible to increase the number of internode gates possible with $10^5$ circuit runs from about 10 to 50. As we highlighted in the previous section these types of infidelities are readily achievable with near-term “on-demand” transducer based links. Quantum links which are enhanced by error mitigation will demonstrate a better scaling potential than classical circuit cutting, somewhat analogous to the recent success in passing the threshold of error correcting codes and demonstrating a favorable scaling there. 

Furthermore, for larger circuits, error mitigation may already be enough to provide a useful advantage to expanding quantum processors with remote quantum links. Bravyi et al. also explored what was possible with optical “t-couplers”~\cite{Bravyi2022}. These microwave-to-optics converter based links would directly multiply the number of qubits available in a processor and provide an extra level of flexibility in the connection of modules which would allow for different connection configurations for use in error mitigation.

\begin{table*}[ht]
    \centering
    \caption{Architecture requirements and advantages}
    \begin{tabular}{p{24mm}|p{50mm}|p{50mm}|p{50mm}}
    Approach & QC limitation & M-to-O requirements & Benefits to QC \\ \hline
    Remote Graph State & T-state generation capabilities with number of physical qubits in fridge & -On-demand entanglement  \newline -Parallelization to 10’s of channels & Reduction in overhead of T-state generation \\ \hline
 Sparse Links   & Local resources: chip-to-chip couplers and/or i/o capabilities	 & -Unknown improvement in the channel entanglement rate or infidelity \newline -OR error mitigation of entanglement links	 & -Quantum volume increased \newline -More resources for algorithm implementation\\ \hline
Lattice Surgery &	Local resources: chip-to-chip couplers and/or i/o capabilities	 & Link gate time improved by 10-100 &		-Full connectivity of all system qubits \newline -Not limited by remote links \\
    \end{tabular} 
    \label{tab:table2}
\end{table*}

\subsection{Remote graph state generation}

An architecture for fault-tolerant computing that requires a distinctly different use of interconnects is that analysed by Saadatmand et al. \cite{Saadatmand2024}. Here they scrutinise a measurement-based approach to algorithm execution. Due to space constraints they consider a monolithic processor containing up to $10^6$ physical qubits arranged in a square lattice with individual dies connected through inter-chip couplers. The algorithm proceeds by preparing and consuming graph states within the monolithic processor, executing a widget (a subcircuit with a maximum circuit depth). Following consumption of the graph state the output is then teleported to a second processor, which has already been prepared in the required graph state (see Figure~\ref{fig:architectures}c). While they consider a brief teleportation time ($\sim$1 $\mu$s for teleportation of a state with the same number of physical qubits as the code-depth of the algorithm), and fault-tolerant teleportation of the output of each subcircuit is required, the architecture is generous to the inter-module communication in two key ways:

\begin{enumerate}
\item The time spent on intra-module operation is ample to prepare and distill the long-range entangled states that enable teleportation between each cycle of graph consumption

\item The number of inter-module physical channels that are required in each ‘pipe’ corresponds to the distance of the code, so links are not required for all edge states such as in lattice surgery. The inter-module pipes are only required to transfer the states of the logical qubits between the widgets.
\end{enumerate}

As such this architecture is particularly well suited to the ‘on-demand’ entanglement generation we consider, where the long-range entangled states are required to be delivered with sufficient fidelity at a particular scheduled time. 

\subsection{Optical switching}

Many hardware implementations for quantum computers rely on fixed geometries which define the neighboring or nearby accessible qubits. Entanglement and other resources can be routed via state transfer through intermediate qubits, but this uses computation time and exposes the transferred states to extra errors. An alternative approach is dynamic reconfiguration during the computation. Using optical switches to change the path of the generated quantum links allows for reconfiguration of the quantum computer during computations (see Figure~\ref{fig:architectures}d).

Optical switches enable large networks of connectivity \cite{Cheng2018}. Widely available MEMS switches can provide all-to-all connectivity at millisecond timescales and state-of-the-art electro-optic switches can be reconfigured at tens of ns timescales, matching the timescales of physical gates for individual qubits. These switches could be deployed for all-to-all connectivity of quantum processors or logical qubits, where the configuration is updated with every logical clock step. 

In the near-term, the main bottleneck towards such widespread connectivity will probably be the capacity of the quantum links. Likely, millions of links will be needed for a full quantum computer. However, given limited link generation hardware, optical switching could be applied to switch entanglement generation capacity to different regions of a quantum computer for the above applications like sparse links and remote graph states, targeting regions where remote resources bring in the most additional benefit.

\subsection{Beyond the surface code}

As we discussed in the previous sections, lattice surgery with two-dimensional error correction codes places particularly strict constraints on the optical links which are generated at the edges. Alternative encodings for the quantum information which is passed through the transduction link can help to reduce these constraints. One example is hyperbolic floquet codes, which have been theoretically demonstrated to reduce link overhead for distributed quantum computation~\cite{Sutcliffe2025}. Indeed, the infidelity for on-demand quantum links using distillation of the examples in section~\ref{subsection:examplelinks} passes the error threshold for hyperbolic floquet codes~\cite{Sutcliffe2025}.

GKP states are emerging as promising candidates for quantum processor components because of their inherent error protection which would in principle reduce the number of physical components required ~\cite{Gottesman2001}. Processors have been built from these states in both the superconducting and the optical domain~\cite{Brady2024, Sivak2023}. A recent theoretical work discovered that GKP states are far less sensitive to the inefficiency of transduction than other states if a GKP state is input into a loss channel of the transducer~\cite{Wang2024}. Propagating optical GKP states as would be required for the input have been demonstrated~\cite{Konno2024}. More work is required to determine the practical requirements of these other encoding schemes, but it is highly likely that some will lead to useful connections via transduction or QFC and optical links between quantum processors employing higher dimensional error correction.

\section{Conclusion and outlook}

As current transducer performance already supports the realization of several advantageous protocols for scaling quantum computers, we expect that the coming years will see the first demonstration of such optical entanglement links between microwave frequency quantum processors in separated cryogenic environments, which will then soon be followed by on-demand links. These links will lead to real benefits to quantum computers through sparse link generation, graph state generation and dynamic optical switching. Given the current expansion and eventual limits to i/o microwave hardware in fridges it is likely that quantum processors with no more than 1,000 to 10,000 physical qubits will need to be supported. This will then consequently require 10-100 on-demand links per cryostat. In order to reach the on-demand threshold and the requisite fidelities, approximately 10-100 transducers per link will therefore be necessary for parallelization or distillation. We hence estimate that in total, quantum computers will require no more than 100-10,000 transducer channels per cryostat, which is in principle achievable with the current technology of integrated, low-power microwave-to-optics transduction.

A few key steps remain for transduction to be able to contribute to the expansion of early quantum computers in real world settings. First, transducers need to implement already attainable improvements to efficiency and noise to allow for on-demand single links with sufficient fidelity. These improvements fall within the limits which have already been demonstrated, but will need to be achieved simultaneously. Furthermore, strategies for improving noise and efficiency have already been demonstrated on similar technological platforms and these can also be applied to microwave-to-optics transducers.

Second, transducers will benefit greatly from integration and co-development with qubit architectures and supporting hardware such as microwave and optical memories to improve performance. While an experimental demonstration of entanglement through an optical link is still in progress, there is currently no known fundamental block to the quantum state transfer process and transfers of weak classical states has already been achieved. Connections with qubits and optical memories will provide a concrete testing bed for pushing the limits of entanglement links with current microwave-to-optics transducers.

Third, in the coming years transducers need to scale up from the presently demonstrated single channel to $>$100 simultaneously operable transducers per dilution refrigerator. This will allow the necessary parallelization for increased rate and distillation for increased fidelity. It will also require supporting technologies such as optical multiplexing and power management in the dilution refrigerator.

We expect that as these three steps are being implemented in practice, theoretical developments in architectures with weaker quantum link requirements will allow these systems to address the biggest roadblocks to quantum computer scaling. Although we do not project that lattice surgery will be possible with existing transducer performances, advances in microwave-to-optics technology have been tremendous in the last decade, and the required performance should become within reach. The ongoing developments in microwave-to-optics transducers technology, including the investigating into a wide array of promising technologies, and the knowledge gained through large-scale transducer production processes will produce the orders of magnitude improvement in efficiency and noise which are required for full scale lattice surgery. 

The fundamental building blocks are in place for the first demonstrations of distributed quantum computation with state-of-the-art processors. It is now time for the constituent technologies such as quantum frequency converters and processors to be connected together and scaled up to applicable sizes. These advances will break quantum processors out of the confines of single units and usher in the first useful real-world applications for quantum computation.

\begin{acknowledgments}
	We would like to thank Mathilde Lemang and Christophe Jurczak for their feedback and detailed comments on the manuscript.
\end{acknowledgments}

\bibliography{References}

\end{document}